\documentclass[]{aa} 
\usepackage[utf8]{inputenc}

\usepackage{graphicx}
\usepackage{siunitx}
\usepackage{amsmath}
\usepackage{textcomp}
\usepackage{multirow}               
\usepackage{booktabs}                              
\usepackage{multicol,lipsum}
\usepackage{colortbl}

\usepackage{txfonts}
\usepackage[]{hyperref}
\usepackage{hyperref}
 \hypersetup{
     colorlinks=true,
     linkcolor=blue,
     filecolor=blue,
     citecolor =blue,      
     urlcolor=blue,
     }

\usepackage{makecell} 

\DeclareRobustCommand{\ion}[2]{%
\relax\ifmmode
\ifx\testbx\f@series
{\mathbf{#1\,\mathsc{#2}}}\else
{\mathrm{#1\,\mathsc{#2}}}\fi
\else\textup{#1\,{\mdseries\textsc{#2}}}%
\fi}
\definecolor{darkred}{rgb}{0.6,0,0}

\usepackage{comment}
\usepackage{balance}
\usepackage{subfig}

\begin{document}

   \title{Minimum observations for accurate Exoplanetary mass estimation with Radial Velocity Analysis ({\tt MINERVA})}

   \titlerunning{MINERVA}
   \author{C. Di Maio
          \inst{\ref{inst:oapa}}\thanks{Corresponding author: \email{claudia.dimaio@inaf.it}}
          \and
           S. Benatti \inst{\ref{inst:oapa}}
           \and
            S. Colombo \inst{\ref{inst:oapa}}
            \and 
            G. Piccinini \inst{\ref{inst:oapa}, \ref{inst:unipa}}
            \and
            R. Spinelli \inst{\ref{inst:oapa}}
            \and 
            V. Fardella \inst{\ref{inst:unipa}, \ref{inst:oapa}}
            \and
            R. Claudi \inst{\ref{oapadova},\ref{ist:rm3}}
            \and
            G. Micela \inst{\ref{inst:oapa}}
        }

   \institute{
    {INAF – Osservatorio Astronomico di Palermo, Piazza del Parlamento, 1, 90134 Palermo, Italy} \label{inst:oapa}
    \and{Università degli Studi di Palermo, Dipartimento di Fisica e Chimica, via Archirafi 36, Palermo, Italy} \label{inst:unipa}
    \and{INAF – Osservatorio Astronomico di Padova, Vicolo dell'Osservatorio 5, 35122 Padova, Italy} \label{oapadova}
      \and{Dipartimento di Matematica e Fisica, Universit\`a Roma Tre, Via della Vasca Navale 84, 00146 Roma, Italy}\label{ist:rm3}
    }

   \date{Received 25 May 2026; Accepted 9 November 2026 }

\abstract{Precise planetary mass measurements are essential for constraining the internal structure and envelope properties of exoplanets. For current and future characterisation programs, such data are crucial to break key degeneracies in atmospheric retrievals. Estimating the observational effort needed to reach the desired mass precision is therefore fundamental for the optimisation of current and future radial velocity (RV) surveys.}
{We present \texttt{MINERVA}, a tool designed to estimate the required number of RV observations to measure transiting planet masses to a given precision, providing a robust scheduling framework for large-scale samples.}
{\texttt{MINERVA} simulates realistic RV time series for targets orbiting FGKM-type stars. Recognising that signal recovery heavily depends on temporal sampling, the code explicitly ingests real-world observational timestamps to rigorously account for weather interruptions and seasonal visibility gaps. To model stellar activity, the pipeline draws ages from local neighbourhood distributions, uses empirical calibrations to derive rotation and magnetic activity levels, and translates these into time-varying RV signals. During development, we also derived a new empirical calibration between chromospheric activity and RV dispersion for K-type stars, using Global Architecture of Planetary Systems (GAPS) and \textit{Ariel} Mass Survey (ArMS) data.}
{We successfully validated \texttt{MINERVA} on targets with known planetary masses, yielding required observational efforts highly consistent literature values. To demonstrate its capabilities on statistically significant samples, we applied the code to a subsample of the \textit{Ariel} Mission Candidate Sample. Adopting HARPS-N/TNG as the reference instrument, we estimated the observing nights necessary to achieve a 30\% planetary mass precision, providing an optimised observational roadmap for atmospheric characterisation.}
{Beyond observing proposal preparation, \texttt{MINERVA}'s probabilistic and flexible architecture allows the optimisation of ongoing RV campaigns, the evaluating of undersampling risks in active stars, and the dynamic guidance of population-wide characterisation strategies.} 


   \maketitle
%
%________________________________________________________________

\section{Introduction}
With over 6000 extrasolar planets identified to date \citep[via the NASA Exoplanet Archive,][]{Christiansen2025PSJ.....6..186C}\footnote{Data retrieved from the NASA Exoplanet Archive, \url{https://exoplanetarchive.ipac.caltech.edu/}}, the field of exoplanetary science is driven by two complementary paths: the statistical characterization of the planet population, continually mapping occurrence rates and architectural diversity since the pioneering eras of the \textit{CoRoT} and \textit{Kepler} space missions \citep[e.g.,][]{Howard2012ApJS..201...15H, Dressing2015ApJ...807...45D, Pinamonti2022A&A...664A..65P}, and the in-depth investigation of individual systems \citep[e.g.,][]{Winn2015ARA&A..53..409W, Carleo2020AJ....160..114C, Naponiello2025A&A...701A..79N}.

In this framework, transiting planets are prime targets as they allow for the direct determination of multiple coupled properties, including radius, bulk density, and atmospheric structure \citep[e.g.,][]{Madhusudhan2019ARA&A..57..617M, Zeng2019PNAS..116.9723Z}. 
However, a comprehensive characterisation of any planetary system inherently relies on the determination of accurate planetary masses \citep[e.g.,][]{DiMaio2023A&A...669A.150D}. 

Obtaining these masses requires observational strategies tailored to the orbital scale. For planets on wide orbits, these RV measurements are increasingly combined with \textit{Gaia} astrometry to constrain the system architectures \citep[e.g.,][]{GaiaCollaboration2016A&A...595A...1G, Piccinini2026A&A...707A.104P}. Conversely, the RV method remains the baseline choice for short- to intermediate-period configurations, as proven by decades of high-precision spectroscopic campaigns \citep[e.g.][]{Knutson2014ApJ...785..126K, Bryan2019AJ....157...52B, Palatnick2021ApJ...909L...6P, Benatti2021A&A...650A..66B}. In these systems, coupling RVs with transit observations breaks the orbital inclination degeneracy \citep[e.g.,][]{Charbonneau2000ApJ...529L..45C, Charbonneau2002ApJ...568..377C}.

Crucially, for these transiting systems, a precise knowledge of the planetary mass is a fundamental prerequisite for robust atmospheric characterisation, the primary focus of the upcoming ESA \textit{Ariel} mission, which will observe approximately 1000 exoplanetary atmospheres \citep{Tinetti2018ExA....46..135T}. Although atmospheric retrieval techniques can theoretically constrain planetary masses directly from transmission spectra \citep[e.g.,][]{deWit2013Sci...342.1473D}, severe parameter degeneracies arise in realistic scenarios. High-altitude clouds in giant planets, or the presence of secondary atmospheres in low-mass planets, introduce direct degeneracies between the main atmospheric components, the planetary radius, and the mass \citep{Batalha2019ApJ...885L..25B, Changeat2020ApJ...896..107C, DiMaio2023A&A...669A.150D}. Prior RV mass measurements with a precision of at least 30\% to 40\% are therefore mandatory to break these degeneracies and fully exploit the unprecedented information content of the \textit{Ariel} spectra \citep{DiMaio2023A&A...669A.150D}.

Recognizing this critical need, the \textit{Ariel} Masses Survey (ArMS, \citealt{Bonomo2026A&A...707A.197B, DiMaio2026A&A...709A.270D}), a five-year Large Programme (PI: S. Benatti) conducted with the HARPS-N spectrograph \citep{Cosentino2014SPIE.9147E..8CC} within the Global Architecture of Planetary Systems (GAPS) collaboration \citep[e.g.,][]{Covino2013A&A...554A..28C, Carleo2020A&A...638A...5C}, was initiated to robustly determine the masses of transitional planets in the radius valley (\citealt{Fulton2017AJ....154..109F, Owen2013ApJ...775..105O}).
However, executing such extensive ground-based follow-up requires accurate operational forecasting. To efficiently plan observational campaigns and optimize telescope time allocation without wasting valuable resources, observers must precisely evaluate the temporal baseline and the number of observations required to achieve a confidence-level detection of the RV semi-amplitude $K$ of at least $3\sigma_{\rm K}$ or $5\sigma_{\rm K}$ (corresponding to a fractional uncertainty $\sigma_{\rm K}/K \le 33\%$ or $\le 20\%$).

This prediction and scheduling analysis can be conducted using several software packages designed for RV modelling \citep[e.g.,][]{Balsalobre-Ruza2023A&A...669A..18B}, where stellar activity is often modelled via Gaussian Processes \citep[][]{2006gpml.book.....R}, although this feature frequently remains optional.
While these standard packages are used for detailed parameter retrieval on individual systems, scaling them up to evaluate entire planetary populations presents two practical challenges. First, exploring parameter spaces sequentially for hundreds of targets introduces a massive computational overhead. Second, they are usually intrinsically data-driven. For instance, forecasting tools such as the \texttt{RV Follow-Up Calculator} \citep{Cloutier2018AJ....156...82C} can model correlated RV noise, but they require high-precision photometric inputs (such as the $\sim$8-hour timescale $F_8$ flicker; \citealt{Cegla2014ApJ...780..104C}) to properly estimate the jitter. Without such pre-existing observations, these algorithms must fall back to a computationally cheaper white-noise approximation. However, assuming purely white noise fundamentally neglects the time-correlated stellar signals, such as granulation, oscillations, and magnetic activity, that could dominate real RV observations \citep[e.g.,][]{Benatti2021A&A...650A..66B, DiMaio2024A&A...683A.239D}. Consequently, this simplification underestimates the true RV scatter, leading to overly optimistic predictions for the required number of measurements, particularly for active or late-type stars.

To address the specific needs of population-level forecasting prior to data collection, we present the \texttt{MINERVA} (MInimum observations for accurate Exoplanetary mass estimation with Radial Velocity Analysis) tool\footnote{\url{https://github.com/EXOPA-OAPa/MINERVA}}. It is an automated, computationally lean code designed to determine the number of observations required to measure the mass of transiting planets orbiting FGKM-type stars. Unlike data-dependent fitting tools that require prior observations for single targets, \texttt{MINERVA} acts as an ab initio population forward-modelling simulator, generating the expected activity-induced RV signal entirely from scratch. It achieves this by treating the targets statistically as part of a synthetic stellar population: fundamental properties (such as mass, age, and rotation) are either drawn from catalogues or sampled from representative demographic distributions. By coupling these physical parameters with empirical scaling laws, the pipeline provides a physically realistic, statistically robust characterization of stellar activity. This allows the code to quickly predict the telescope time needed to across massive lists without the severe computational bottlenecks of Bayesian retrievals.
By enabling reliable a priori forecasting, \texttt{MINERVA} is exceptionally valuable for planning extensive follow-up campaigns for missions like PLATO \citep{Rauer2014ExA....38..249R, Rauer2025ExA....59...26R} and \textit{Ariel}, which aim to characterise hundreds to thousands of transiting planets.

In this paper, we apply \texttt{MINERVA} to a representative subsample of the \textit{Ariel} Mission Candidate Sample (MCS, \citealt{Edwards2022AJ....164...15E}), released in July 2025, from which the final Mission Reference Sample (MRS) \footnote{\url{https://github.com/arielmission-space/Mission_Candidate_Sample}} will be extracted. Using HARPS-N at the Telescopio Nazionale Galileo (TNG) as the reference instrument, we evaluate the total observing time necessary to fully characterize this sample.

The paper is organized as follows: Sect. \ref{sec:setup} details the complete \texttt{MINERVA} architecture, from the simulation of stellar activity, to the injection of the planetary signal, the definition of the observational cadence, and the signal recovery process. Section \ref{sec:results} presents our results: first, we validate the code's reliability against published literature (Sect. \ref{sec:validation}), and subsequently, we estimate the total telescope time needed for the \textit{Ariel} core survey (Sect. \ref{sec:arielsample}). Finally, conclusions are drawn in Sect. \ref{sec:conclusions}.

\section{MINERVA setup}
\label{sec:setup}
In this section, we describe the elements that constitute the code and the workflow of the simulations. A flowchart of the entire code is reported in Fig.~\ref{fig:diagramma}. \texttt{MINERVA} has been developed in a modular way to ensure flexibility, allowing the user to modify its components so that, for example, it can be applied independently of the instrument or the adopted temporal sampling. However, for the analysis of our sample presented in the following sections, we assume HARPS-N as the reference instrument and adopt the default time samplings described below, which reflect realistic cadences for current high-precision RV programs.

\begin{figure*}[h!]
	\centering
	\includegraphics[width=\textwidth]{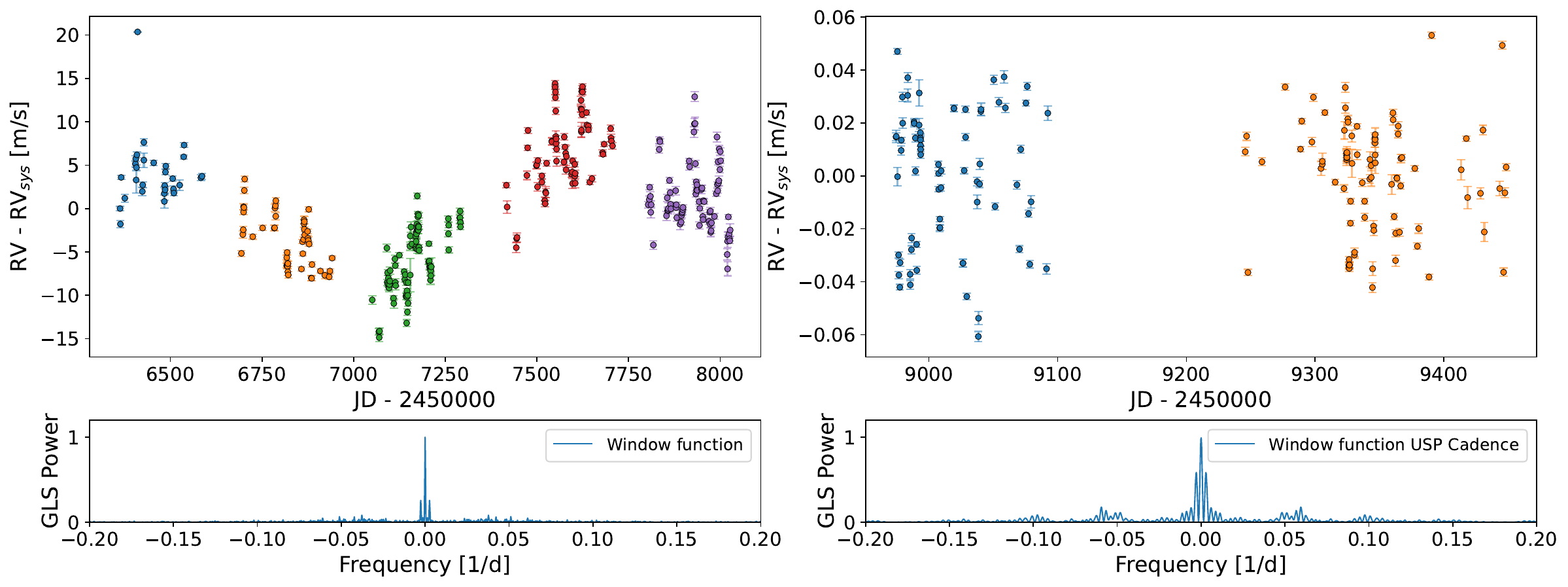}
	\caption{Spectroscopic datasets and their temporal sampling used as default time sampling in this work. The panels illustrate RV time series and corresponding window functions for two different targets.
(Left panels) RV monitoring of HD 164922 performed with HARPS-N, as described by \citet{Benatti2020A&A...639A..50B}.
(Right panels) HARPS-N observations of the ultra-short period planet TOI-1807, presented by\citet{Nardiello2022A&A...664A.163N}.
The datasets are divided into different observing seasons, highlighted by points with distinct colours.
The bottom panels display the window functions computed from the {\tt GLS} periodogram.}
		\label{fig:harps_timesampling}
\end{figure*}
\subsection{Definition of the time coordinate}\label{subsec:time}
To obtain realistic synthetic RV time series, it is necessary to  simulate the contributions of the star, the planet and the instrument. 
The first step in this process is defining the time baseline of the synthetic observations. To realistically replicate the seasonal visibility of the targets on the sky and the subsequent periods of non-visibility, we simulate a three-year total baseline composed of alternating six-month observing seasons and six-month visibility gaps. Within each active six-month observing window, we generate 60 time stamps randomly distributed to mimic realistic, non-uniform scheduling constraints. This setup results in a total of 180 epochs across the entire three-year observational campaign. Clearly, this approach represents an idealised case that is rarely achievable under real observing conditions. 
Initial benchmarking tests performed on simulated quiescent stars ($\log R'_{\rm HK} < -5.0$) demonstrated that an idealized, nearly uniform random distribution of observations artificially reduces the number of data points required for a robust planet detection compared to actual, ground-based observing campaigns. We conducted a controlled experiment holding the underlying noise properties (both instrumental white noise and stellar jitter variance) strictly identical, while varying exclusively the time stamps of the observations. The idealized sampling minimizes the sidebands of the spectral window function, concentrating the planetary signal power into a single clean peak in the periodogram. Conversely, realistic cadences which includes weather interruptions, scheduling conflicts, and daily or monthly gaps, introduce severe aliasing and power leakage. This power degradation requires a larger number of epochs to achieve the same statistical significance and parameters accuracy, making the adoption of realistic seasonal architectures mandatory for the predictive goals of this work.

For this reason, we adopt realistic distributions of time stamps to obtain reliable results. While the \texttt{MINERVA} pipeline is designed to be fully flexible and can ingest any arbitrary, user-defined time series, for the automated statistical simulations presented throughout this work we implement two standardized default presets tailored to the orbital period of the planet: a ``typical'' distributed tracking scenario and an ``extreme'' high-density monitoring scenario. The ``typical'' strategy simulates a standard, long-baseline observing program where a target is sampled two or three times per week throughout its six-month visibility window, naturally accounting for routine weather interruptions and technical telescope downtimes. Conversely, the ``extreme'' strategy represents a highly concentrated, ultra-high-cadence campaign (e.g., multiple observations per night over consecutive allocation blocks). This ``extreme'' scenario denotes an exceptionally dense cluster of observations, specifically chosen to optimize the phase coverage of ultra-short-period (USP) planets or to resolve high-frequency stellar activity variations.
For targets with orbital periods $P$ $\ge$ 2 days, our default sampling is based on the high-cadence strategy used for the target HD 164922 within the GAPS program \citep{Benatti2020A&A...639A..50B}. This consists of 300 time frames collected over 6 years. In the first and second observing seasons, 76 RV data points were obtained over 19 months, while the third, fourth, and fifth seasons feature particularly dense sampling. 

While this baseline strategy is robust for most targets, it may be insufficient for USP planets ($P < 2$ days). \citet{Murphy2025AJ....169...25M} tested the impact of observing cadence on the accuracy of small planet mass measurements, recommending the acquisition of 2--3 measurements per orbit of the innermost planet and a minimum of 40 RVs epochs. Following these recommendations, for planets with $P < 2$ days we adopt a second default strategy specifically designed for intensive monitoring.
For this purpose, we used the temporal sampling of TOI-1807, which comprises 160 RV epochs obtained over two years within the GAPS program \citep{Nardiello2022A&A...664A.163N}. In this case, the observing strategy was specifically optimised for USP planets, featuring a cadence of at least two RV observations per night.

These data cadences  realistically reflect the scheduling of a large RV program, accounting for target visibility and gaps due to weather constraints. These observing strategies represent particularly favourable cases, achievable only under optimal and sustained operational conditions, and remains above the cadence typically attainable for most targets. Figure \ref{fig:harps_timesampling} shows the complete time distribution for both cases along with the associated window function.

\subsection{Definition of the RV noise floor and measurement uncertainties} \label{subsec:noise}

To simulate the dispersion of the RV time series, we generated an intrinsic noise component using a Gaussian distribution. The total dispersion of this distribution, $\sigma$, is obtained by combining several independent physical and instrumental contributions. Following the scheme described by \citet{Cloutier2018AJ....156...82C}, we defined:
\begin{itemize}
    \item[-] Instrumental noise, $\rm \sigma_{\rm RV,instr}$: in the case of HARPS-N at TNG, this is estimated to be 0.5\,m\,s$^{-1}$ \citep{Cosentino2014SPIE.9147E..8CC}; 
    \item[-] Photon noise, $\rm \sigma_{\rm RV,ph}$: estimated for the HARPS-N mode under given observing conditions using the exposure time calculator included in the New Short Term Scheduler\footnote{Available via the TNG web page \url{https://www.tng.iac.es/instruments/harps/tools/NSTS.jar}};
   \item[-] Contribution from additional planets in the system, $\rm \sigma_{\rm RV,pl}$: fixed at 2\,m\,s$^{-1}$ to account for the dynamical unresolved background of unknown multiple planets. This value acts as a deliberate, conservative noise floor representing unresolved planetary jitter. Statistically, multi-planet systems are ubiquitous, with super-Earths and sub-Neptunes being the most common companions \citep[e.g.,][]{Mayor2011arXiv1109.2497M, Fressin2013ApJ...766...81F, Petigura2013PNAS..11019273P}. From a data-analysis perspective, much larger companions would be quickly resolved and modelled out of the RV time series, while much smaller terrestrial planets would fall negligibly below the instrumental noise. An amplitude of 2\,m\,s$^{-1}$ represents the typical intermediate regime that effectively acts as correlated noise, hindering the mass characterisation of the primary target. Adopting this standard margin prevents the pipeline from simulating unrealistically clean datasets and underestimating the total observational effort.
\end{itemize}

All three noise terms are implemented in the code as user-defined parameters, allowing full flexibility in adapting the simulations to different instruments, observing strategies, or scientific goals. The effective RV scatter injected into the simulated time series is thus given by 
\begin{equation}
\rm \sigma = \sqrt{\sigma_{\rm RV,instr}^2 + \sigma_{\rm RV,ph}^2 + \sigma_{\rm RV,pl}^2}.
\end{equation}
It is important to note that the correlated stellar activity signal is simulated separately (as detailed in Sect. \ref{subsec:activity_sim}) and directly added to the  planetary signal and white noise component.

While $\sigma$ dictates the true underlying physical scatter of the mock data points, it does not represent the formal uncertainty known to the observer. In active targets, stellar activity distorts the spectral lines (e.g., affecting line symmetry and depth), directly hindering the numerical precision of the radial velocity extraction algorithms, such as the standard Gaussian fit to the cross-correlation function profile \citep[e.g.,][]{DiMaio2024A&A...683A.239D}. To mimic real data reduction procedures, we assign to each $i$-th simulated epoch a formal measurement uncertainty $\delta RV_i$. This error bar is calculated by summing in quadrature the nominal instrumental noise, photon noise, and an activity-driven extraction term ($\sigma_{\rm act, extr}$):

\begin{equation}
\delta RV_{\rm i} = \sqrt{\sigma_{\rm \text{RV,instr}}^2 + \sigma_{\rm \text{RV,ph}}^2 + \sigma_{\rm \text{act, extr}}^2}.
\end{equation}
This approach ensures that the statistical weights used during the periodogram analysis and the RV fitting process (detailed in Sec. \ref{subsec:recovery}) reflect the physical limits of data reduction pipelines when processing active targets.

To assign realistic values to $\sigma_{\rm \text{act, extr}}$, we exploited an extensive database of observations collected within the framework of the GAPS programme. The reference sample consists of 250 FGKM main-sequence stars, with a total of 15\,270 observations,  spanning a wide range of magnetic activity levels. The parameter space covers values of the $\log R'_{\rm HK}$, a standard proxy for stellar chromospheric activity derived from the Ca~{\sc ii} H\& K emission lines, from --6.0 to --3.6. This interval corresponds to empirical RV jitter amplitudes due to the stellar activity from less than 1\,$\rm m\,s^{-1}$ for the quietest targets up to about 100\,$\rm m\,s^{-1}$ for the most active cases. For late-K and M dwarfs, the $\log R'_{\rm HK}$ values were derived from the measured $S$-index using the empirical calibrations by \citet{SuarezMascareno2015MNRAS.452.2745S} for K stars with $1.2 < B-V \le 1.7$\,mag and M stars with $B-V \le 1.7$\,mag, and by \citet{Astudillo-Defru2017A&A...600A..13A} for the reddest targets with $B-V > 1.7$\,mag.

Based on this database, \texttt{MINERVA} assigns an empirical extraction degradation value ($\sigma_{\rm \text{act, extr}}$) to each mock time series by matching the simulated star's spectral class and assigned activity level ($\log R'_{\rm HK}$) against the real target distribution. A comprehensive description of the database and the empirical relations between stellar activity and raw pipeline extraction uncertainties will be presented in a dedicated work (R. Claudi et al., in prep.).
Specifically, the pipeline first filters the database by spectral class (F, G, K, or M) so that a simulated target only draws extraction errors from real stars of the same spectral type. Once the spectral class is fixed, the code defines a local activity window by restricting its search to a narrow interval centred on the mock star's assigned activity level (spanning $\log R'_{\rm HK} \pm 0.1~\text{dex}$), ensuring that the sampled extraction uncertainty corresponds strictly to stars exhibiting the same chromospheric activity regime. From this window, a nominal extraction uncertainty value is randomly drawn and perturbed by a 10\% Gaussian scatter to realistically simulate nightly observational fluctuations caused by atmospheric seeing and airmass variations. This approach guarantees that the final $\delta RV_{\rm i}$ error bars vary from epoch to epoch, accurately mirroring real observational datasets.

\subsection{Stellar population properties}\label{subsection:starpop_properties}

As part of the \texttt{MINERVA} simulation framework, the physical properties and the expected level of magnetic activity of each target star must be defined, as they directly determine the amplitude of the macroscopic RV activity signal ($K_{\rm act}$) injected into the simulated time series. To achieve this, the pipeline is designed with a dual approach: it can either ingest known parameters from specific target catalogues (operating via uncertainty bootstrapping) or autonomously model a synthetic stellar population representative of the solar neighbourhood.

In the unconstrained population mode, starting from the assumption that the sample reflects the typical main-sequence population of the solar neighbourhood, we adopt a constant stellar birthrate and assign to each object a set of physical parameters, such as age, mass, and rotation, consistent with empirical and theoretical relations. These fundamental quantities are then propagated through a chain of scaling laws to derive the chromospheric activity index ($\log R'_{\rm HK}$) and the corresponding RV dispersion.

While we implemented a specific set of well-established empirical relations (detailed below) as our baseline framework to ensure physical continuity across the FGKM spectral sequence, we acknowledge the existence of alternative robust calibrations in the literature connecting stellar rotation, activity, and RV variability \citep[e.g.,][]{SuarezMascareno2016A&A...595A..12S, SuarezMascareno2018A&A...612A..89S}. Thanks to \texttt{MINERVA}'s modular Python architecture, any of these internal scaling laws can be easily modified or completely replaced by the end-user to suit different stellar samples or specific scientific goals.

As a first step, we generated a uniform distribution of stellar ages for each star in the sample, ranging between $10^8$ and $10^{10}$\,yr for stars with masses lower or equal than that of the Sun. For more massive stars, we adjusted the upper limit according to the maximum ages derived from the YAPSI evolutionary tracks \citep{Spada2017ApJ...838..161S}.

The dependence of stellar rotation on age is not unique but varies depending on the regime in which the star evolved after dissipating its circumstellar disk. Roughly, stars may begin their evolution towards the main sequence in one of three regimes: slow, median, or fast rotators \citep{Gallet_Bouvier2015A&A...577A..98G}. After 1–3\,Gyr, stellar rotation converges towards a common track. To simulate this behaviour, we assigned each star a rotational regime by randomly sampling a uniform distribution that reproduces approximately 25\% slow, 50\% median, and 25\% fast rotators. Slow, median, and fast rotators correspond to the typical rotation rates in each category. This assignment is used solely to define the rotational regime and does not directly enter subsequent calculations. The definition of the rotation rate can be adjusted if needed.

Using the sets of stellar ages and assigned rotation regimes, we applied the calibration relations from \citet{Gallet_Bouvier2015A&A...577A..98G} to compute the $P_{\rm rot}$ of each star. These authors modelled the evolution of stellar angular momentum as a function of age and rotation regime for solar-like stars from 1\,Myr to 10\,Gyr. From their work, we extracted sets of coefficients ($A$ and $B$) for three typical stellar masses and the three rotation regimes, applying the relation:
\begin{equation}
\log P_{\rm rot} = A + B \times \log(\rm age)
\label{eq:coefficient_gallet}
\end{equation}
to determine $P_{\rm rot}$ for each star.
The $A$ and $B$ coefficients adopted to compute the stellar rotation period are listed in Table \ref{tab:coeff_Gallet_Bouvier}.

\begin{table}\footnotesize						
\centering										
\caption{Adopted coefficients for Eq. \ref{eq:coefficient_gallet}. }										
\label{tab:coeff_Gallet_Bouvier}
\begin{tabular}{cclc}										
\hline										
\hline										
\rule{0pt}{2.8ex}A & B & Regime & Age (yr) \\										
\hline										
\multicolumn{4}{l}{\normalsize \rule{0pt}{2.8ex}$M$ = 1.0 M$_\odot$} \\[0.6ex]			
-12.99 & 1.574 & Fast   & $<10^9$ \\
-3.12  & 0.477 & Median & $<10^9$ \\
-1.28  & 0.273 & Slow   & $<10^9$ \\ 
-3.53  & 0.523 &  All     & $>10^9$ \\
				
\hline	
\multicolumn{4}{l}{\normalsize \rule{0pt}{2.8ex}$M$ = 0.8 M$_\odot$} \\[0.6ex]		-10.63 & 1.301 & Fast   & $<10^9$ \\
-3.22  & 0.477 & Median & $<10^9$ \\
-1.03  & 0.234 & Slow   & $<10^9$ \\ 
-3.10  & 0.465 &  All     & $>10^9$ \\

\hline	
\multicolumn{4}{l}{\normalsize \rule{0pt}{2.8ex}$M$ = 0.5 M$_\odot$} \\[0.6ex]		-11.43 & 1.354 & Fast   & $< 3 \times 10^9$ \\
-6.44  & 0.827 & Median & $< 3 \times 10^9$ \\
-3.71  & 0.539 & Slow   & $< 3 \times 10^9$ \\ 
-1.31  & 0.301 &  All     & $> 3 \times 10^9$ \\
\hline
\end{tabular}																	
\vspace{0.5em}																
\end{table}																		

To determine the amplitude of the activity contribution to the RV time series, we first derived the convective turnover time ($\tau$) for each star based on its stellar mass and age, using the YAPSI \footnote{\url{http://www.astro.yale.edu/yapsi/} }evolutionary tracks assuming a solar configuration \citep{Spada2017ApJ...838..161S}. Using the previously computed rotation period and $\tau$, we calculated the Rossby number (${Ro = P_{\rm rot}/\tau}$). We then estimated the $ \log R'_{\rm HK}$ as a function of the Rossby number using the relations from \citet{Mamajek2008ApJ...687.1264M} for FGK stars, and we adopted the relation between $ \log R'_{\rm HK}$ and the rotational period from \citet{Astudillo-Defru2017A&A...600A..13A} for M dwarfs.

The RV dispersion induced by stellar activity was then estimated using the relations from \citet{Santos2000A&A...361..265S} for FGK stars, while for M dwarfs we adopted the empirical RV dispersion relation derived from the GAPS program database (L. Affer, priv. comm.).
\begin{figure}[h!]
	\centering
	\includegraphics[scale=0.48]{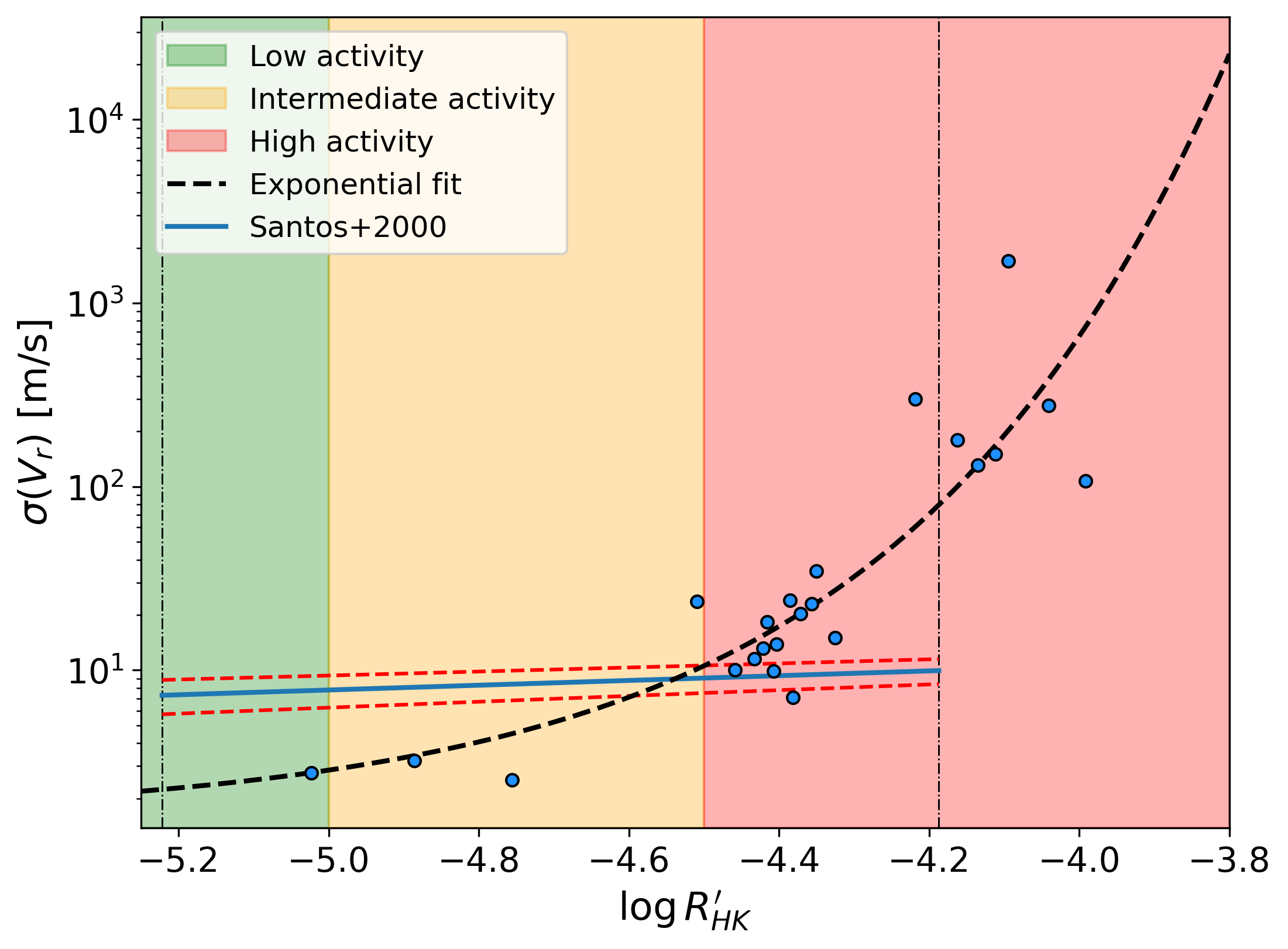}
	\caption{ Reduced radial velocity dispersion, $\sigma(V_{\rm r})$, as a function of $\log R'_{\rm HK}$ for K-type stars in the GAPS subsample. The black dashed line shows the best-fit exponential relation derived from the data. The solid blue line and the dashed red lines indicate the empirical relation and its corresponding rms scatter proposed by \citet{Santos2000A&A...361..265S}, respectively. Notably, these literature curves are plotted exclusively within the activity range explored in their original paper, highlighted by the vertical dot-dashed lines. The background colours highlight different stellar activity level regimes based on $\log R'_{\rm HK}$: low activity (green shaded area, $\log R'_{\rm HK} < -5$), intermediate activity (orange shaded area, $-5 \leq \log R'_{\rm HK} < -4.5$), and high activity (red shaded area, $\log R'_{\rm HK} \geq -4.5$). The two clear outliers in the high and very high activity regimes correspond to V1298 Tau and TAP 26, respectively, which host one or more planets whose signals could not be removed due to the lack of accurate planetary mass estimates.}
\label{fig:newSantos}
\end{figure} 
While the empirical relation proposed by \citet{Santos2000A&A...361..265S} describes the behaviour of mature, relatively quiet field stars, fulfilling the broader scientific goals of the \texttt{MINERVA} simulator required extending the calibration framework for K-type dwarfs to significantly higher activity levels. As the authors themselves noted, their original K-star dataset did not probe extreme activity regimes ($\log R'_{\rm HK} > -4.2$). Due to gyrochronology, such high and very high activity levels are physically tied to early evolutionary stages (i.e., young stars), as mature K dwarfs cannot sustain such enhanced magnetic dynamos.

Historically, young active stars have been largely excluded from intensive RV planet-search campaigns because their pronounced stellar jitter severely hampers the detection of Keplerian signals. Consequently, high-cadence RV datasets for these extreme regimes are relatively scarce in the literature. However, these systems are now recognized as paramount targets for investigating the early stages of planetary formation and evolution. Indeed, several well-studied young K dwarfs display substantial RV variability driven by magnetic activity (e.g., V1298~Tau with $\log R'_\mathrm{HK} > -4.5$ and $\mathrm{rms}_\mathrm{RV} > 100\,\mathrm{m\,s^{-1}}$: \citealp{DiMaio2024A&A...683A.239D, Damasso2026arXiv260930038D, Guilluy2026arXiv261032815G}; and DS Tuc A with $\log R'_\mathrm{HK} > -4.2$ and $\mathrm{rms}_\mathrm{RV} > 160\,\mathrm{m\,s^{-1}}$: \citealp{Benatti2021A&A...650A..66B})

To bridge this gap and provide a continuous operational scaling law across the entire activity space, we derived a new empirical relation for K-type stars using a sub-sample of 25 stars from the GAPS and ArMS programs, spanning values of $\log R'_{\rm HK}$ from $-5.02$ to $-3.99$. Some of these stars host one or more planets, and in such cases, the RV dispersion would be artificially inflated by Keplerian signals; where possible, we removed the planetary contribution using literature orbital parameters and recomputed the RV dispersion to isolate the pure stellar activity component. 
 
Figure \ref{fig:newSantos} shows the reduced radial velocity rms, $\sigma'(V_{\rm r})$, as a function of the $\log R'_{\rm HK}$. For each target, the $\sigma'(V_{\rm r})$ values were computed by subtracting in quadrature the contribution of the individual radial velocity uncertainties from the root mean square of the measured radial velocities. 

Our empirical data outline a clear trend between $\sigma'(V_{\rm r})$ and stellar activity for K-type stars, exhibiting a distinct non-linear behaviour. Within the intermediate activity range shared with the original study ($\log R'_{\rm HK} \lesssim -4.2$), where our sub-sample contains a comparable number of targets, our relation systematically traces a higher scatter and a slightly steeper slope. However, the prominent exponential growth of the RV dispersion occurs exclusively at higher activity levels ($\log R'_{\rm HK} > -4.2$), characterizing an extreme regime.
The final best-fit exponential relation adopted in our pipeline is:
\begin{equation}
\sigma'(V_{\rm r}) = 1.558 \times e^{0.605 R_5},
\end{equation}
where $R_5 = 10^5 R'_{\rm HK}$.

At the end of the procedure described above, each target in our synthetic sample was assigned a predicted radial velocity amplitude due to the predicted stellar activity ($K_{\rm act}$) and rotational period. Although the approach is valid in a statistical sense rather than for individual objects, each step can be easily adapted to model stellar populations with specific characteristics. 

To propagate physical uncertainties and noise variations across an observational campaign, \texttt{MINERVA} executes this parameter derivation framework over $N$ independent simulated realizations (user-defined, set to $N = 100$ by default) for each target system. Across these realizations, the pipeline operates in two distinct modes depending on the available target constraints. In the unconstrained population mode, each iteration draws a stellar age from population distributions and applies the chain of empirical calibrations described above to dynamically derive $P_{\mathrm{rot}}$, $\tau$, $\log R'_{HK}$, and $K_{\mathrm{act}}$, effectively acting as a Monte Carlo sampling engine over stellar evolution and activity models. Conversely, for targets with constrained literature parameters (such as our validation sample), \texttt{MINERVA} features an optional stellar parameter bootstrapping module. When enabled, each iteration resamples key physical properties, such as stellar mass, effective temperature, and $\log R'_{HK}$, directly from Gaussian distributions defined by their reported literature uncertainties ($\pm\sigma$) prior to calculating the activity contributions. In both operational modes, this multi-realization architecture ensures that \texttt{MINERVA} yields realistic distributions of potential observational outcomes rather than a single deterministic prediction.

\subsection{Definition of the stellar activity signal} \label{subsec:activity_sim}
The contribution to the RV dispersion due to varying levels of stellar activity was added to the previously generated pure noise time series. Real active regions (spots and faculae) evolve over time, appear at different latitudes, and have finite lifetimes. To account for these complexities, the stellar activity signal is dynamically generated by incorporating variability that could mimic effects like differential rotation, amplitude evolution, and phase incoherence. 

We categorised the simulated targets stars into three activity regimes based on their assigned $\log R'_{\rm HK}$ values. These boundaries serve as generic operational thresholds to evaluate the pipeline under different noise conditions, reflecting both the empirical bimodal distribution of chromospheric activity and its correlation with radial velocity jitter \citep[e.g.,][]{Henry1996AJ....111..439H, Mamajek2008ApJ...687.1264M, Isaacson2010ApJ...725..875I, Lovis2011arXiv1107.5325L}:

\begin{itemize}
    \setlength{\itemsep}{4pt}
    \item[-] Low activity: $\log R'_{\rm HK} < -5.0$
    \item[-] Intermediate activity: $-5.0 \le \log R'_{\rm HK} < -4.5$
    \item[-] High activity: $\log R'_{\rm HK} \ge -4.5$
\end{itemize}
The baseline properties of the injected signals, namely the primary rotation period $P_{\rm rot}$ and the fundamental activity amplitude $K_{\rm act}$, are derived for each target as detailed in Sect. \ref{subsection:starpop_properties}. 

To simulate the effect of active regions located at different stellar latitudes rotating at different velocities, we introduced a secondary periodic signal $P_{\rm rot,2} = P_{\rm rot}(1 + \alpha)$. Following the approach of \citet{Reinhold2013A&A...560A...4R}, we constrained the differential rotation shear parameter $\alpha$ within the physically motivated boundaries of $[0.01, 0.30]$, where the upper limit is a conservative choice slightly above the solar value ($\alpha_{\odot} \approx 0.20$). Furthermore, empirical observations indicate that fractional shear decreases with increasing rotation rate and magnetic activity \citep[e.g.,][]{Barnes2005MNRAS.357L...1B}. We therefore tied the specific simulated $\alpha$ range to the stellar activity level. Specifically, $\alpha$ was drawn from a uniform distribution $\alpha \in [0.01, 0.08]$ for highly active stars ($\log R'_{\rm HK} > -4.5$), and $\alpha \in [0.10, 0.20]$ for more quiescent targets, consistent with the solar shear. The superposition of these closely spaced frequencies creates a beating pattern that naturally modulates the amplitude and shape of the RV curve.
The complexity of the injected signal dynamically adapts to the activity regime.

For the low-activity case, we assume the star hosts stable, long-lived active regions. The signal is modelled as a continuous, phase-coherent sum of the primary rotational modulation (with an amplitude $K_{\rm act}$) and a minor secondary disturbance at $P_{\rm rot,2}$ set to 10\% of $K_{\rm act}$, preventing the signal from being a mathematically perfect, unnatural sinusoid.
For any observation time $t$, this is expressed as:
\begin{equation}
    RV_{\rm act}(t) = K_{\rm act}\sin\bigg(\dfrac{2\pi}{P_{\rm rot}}t + \phi_1\bigg) + 0.1K_{\rm act}\sin\bigg({\dfrac{2\pi}{P_{\rm rot,2}}t+\phi_2}\bigg)
\end{equation}
where $\phi_1$ and $\phi_2$ are fixed global phases.

For the intermediate- and high-activity cases, the assumption of long-term phase coherence breaks down due to the finite lifetime of starspots and active complexes, which typically ranges from a few weeks to a few months \citep[e.g.,][]{Lanza2009A&A...493..193L, Dumusque2011A&A...527A..82D, Giles2017MNRAS.472.1618G}. To capture this physical evolution within our observational framework, the simulated time series is divided into independent observing seasons. Because the seasonal gap is longer than the typical decay timescale of individual active regions, the stellar surface topology undergoes a complete reconfiguration from one year to the next. This naturally justifies performing macrostructural resets at seasonal boundaries rather than within a single continuous observing window. For each season $s$, the algorithm simulates the emergence of new active regions through three mechanisms:
\begin{enumerate}
    \item {Phase resets:} The fundamental and secondary signals are assigned new, independent random phases $\phi_{\rm s,n} \sim \mathcal{U}(0,2\pi)$ at the transition of each season to simulate the decay of old spots and the emergence of new ones.
    \item {Amplitude evolution:} The base amplitude $K_{\rm s}$ for a given season is drawn from a normal distribution $\mathcal{N}(K_{\rm act}, 0.1 K_{\rm act})$ to simulate variations in spot coverage. This value is bounded to a strict minimum of 0.5\,$\rm m\,s^{-1}$, a threshold consistent with the intrinsic radial velocity floor induced by the stellar granulation and supergranulation in late-type stars \citep[e.g.,][]{Dumusque2011A&A...525A.140D, Meunier2015A&A...583A.118M, Dalal2023MNRAS.525.3344D}.
    \item {Harmonic complexity and strong beating:} Continuous, short-term modifications of the RV curve within a single observing season are naturally driven by the differential rotation beating pattern ($P_{\text{rot}}$ vs $P_{\text{rot},2}$), eliminating the need for abrupt phase discontinuities inside a continuous observational block. The analytical formulation adapts to the specific activity level. 
    For intermediate-activity stars, the amplitude ratio between the secondary and primary signals is set to 0.4. This mimics a stellar surface dominated by one primary active latitude, with a secondary, weaker spot group generating an amplitude envelope modulation:
    \begin{equation}
    RV_{\rm act}(t) = K_{\rm s}\sin\bigg(\dfrac{2\pi}{P_{\rm rot}}t + \phi_{s,1}\bigg) + 0.4K_{\rm s}\sin\bigg({\dfrac{2\pi}{P_{\rm rot,2}}t+\phi_{s,2}}\bigg).
\end{equation} 
For high-activity stars, this ratio is increased to 0.8 to create highly irregular beating patterns. Furthermore, in the high-activity regime, the first harmonic ($P_{\rm rot}/2$) is added with an amplitude of $0.3 K_s$ to reproduce the complex, non-sinusoidal shapes caused by multiple spot groups distributed across opposite stellar hemispheres: 
\begin{equation}
\begin{split}
    RV_{\rm act}(t) = {} & K_{\rm s}\sin\bigg(\dfrac{2\pi}{P_{\rm rot}}t + \phi_{s,1}\bigg) + 0.8K_{\rm s}\sin\bigg(\dfrac{2\pi}{P_{\rm rot,2}}t+\phi_{s,2}\bigg) \\
    & + 0.3K_{\rm s}\sin\bigg(\dfrac{4\pi}{P_{\rm rot}}t + \phi_{s,3}\bigg),
\end{split}
\end{equation}
where $\phi_{s,3}\sim\mathcal{U}(0,2\pi)$ represents the independent seasonal phase of the harmonic component. 
\end{enumerate}
In this way, the final activity time series captures both the stochastic and quasi-periodic nature of stellar activity, preserving the underlying rotational modulation while realistically mimicking the temporal evolution of the stellar photosphere.

\subsection{Definition of the planetary signals}\label{subsec:planetary_signal}

Planetary signals are injected into the mock time series as Keplerian radial velocity variations. For a circular orbit ($e=0$), this reduces to a pure sinusoidal function with a theoretical semi-amplitude ($K_{\rm p}$) defined as \citep{lovis2010radial}:
\begin{equation}
    K_{\rm p} = 28.43\ \mathrm{\,m\,s^{-1}} \left( \dfrac{m_{\rm p}\sin i}{M_{\rm Jup}} \right) \left( \dfrac{m_{\rm p} + m_{\star}}{M_\odot}\right)^{-2/3} \left( \dfrac{P}{\rm yr}\right)^{-1/3},
\end{equation} 
where $m_{\rm p}$ is the planet mass, $i$ is the orbital inclination relative to the sky plane, and $m_{\star}$ is the stellar mass. 

It is important to note that \texttt{MINERVA}'s fast signal-recovery module (described in Sect.~\ref{subsec:recovery}) intentionally relies on circular orbital fits. This design choice mirrors standard blind RV follow-up strategies, where the orbital eccentricity of a newly discovered transiting candidate is generally assumed to be zero prior to the mass measurement. However, the simulation engine (forward-modelling) includes the option for the end-user to manually define non-zero eccentricities ($e$) and arguments of periastron ($\omega$). This deliberate decoupling allows users to explicitly investigate the impact of unmodeled eccentricity on the retrieval efficiency, for instance, quantifying how many additional RV epochs are required to robustly detect an eccentric planet using a circular-orbit scheduling baseline.

For the baseline application to the \textit{Ariel} sample presented in this work, we simulated and retrieved purely circular orbits ($e=0$). This is a highly reasonable approximation given that most planets in the \textit{Ariel} MCS have very tight orbits, and their inclination is constrained to a transiting configuration ($i \sim 90^\circ$). Furthermore, an analysis of the NASA Exoplanet Archive \citep{Akeson2013PASP..125..989A} shows that, among planets with measured eccentricities and orbital periods, about 84\% have $P_{\rm orb} < 50$ days and $e < 0.07$, strongly supporting the circular orbit assumption for this specific population. 
When the planetary mass is not directly measured, it is estimated from the mass–radius relation of \citet{Chen_Kipping2017ApJ...834...17C}, as implemented by \citet{Louie2018PASP..130d4401L}.

%\begin{figure*}[h!]
%	\centering
%	\includegraphics[width=\textwidth]{img/K2-370_timeseries.pdf}
%	\caption{Top panels: RV time series generated by \texttt{MINERVA} for K2-370b, including the instrumental noise floor, the stellar activity contribution, and the planetary signal. The left panel shows 100 data points randomly selected from the full set of 300 simulated observations, while the right panel displays 222 points, the mean minimum number of RV measurements required to achieve a significant detection of the planetary signal, averaged over 100 simulation runs by \texttt{MINERVA}. Different colours indicate the subsets of observations progressively added during the simulation process.
%    Middle panels: Generalised Lomb–Scargle (GLS) periodograms computed from the corresponding RV time series. The red point marks the highest peak, corresponding in this case to the stellar rotation period, while the red dashed horizontal lines indicates the false-alarm probability (FAP) levels at 0.1\% and 1\%. The green vertical line shows the orbital period of the planet. 
%    Bottom panels: GLS periodograms of the residuals obtained after removing the most significant periodicity. The red point marks the highest peak in the residuals, and the red dashed horizontal lines indicate the FAP levels at 0.1\% and 1\%. The green vertical line again shows the planetary orbital period.}
%\label{fig:K2-370}
%\end{figure*} 

\subsection{Estimation of the number of RV measurements required for signal recovery}\label{subsec:recovery}
To determine the number of RV data points required to fully recover the injected planetary signal, we simulate a realistic observing campaign using an incremental approach. We start by selecting an initial set of 10 data points and then add 5 subsequent RV measurements at a time. Following each addition, the pipeline performs a sequence of periodogram analyses and formal modelling to assess the signal recovery. 
The iterative loop terminates as soon as all recovery criteria are met, yielding the required $N_{\mathrm{RV}}$ threshold for that specific realization. It is important to clarify that while this value represents the minimum number of data points needed by our automated fitting framework to statistically claim a detection, the resulting $N_{\mathrm{RV}}$ serves as a deliberately conservative upper limit for survey planning purposes, as the pipeline does not apply advanced post-processing activity mitigation techniques (such as Gaussian process). 

Initially, a Generalized Lomb-Scargle (GLS) periodogram \citep{Zechmeister2009A&A...496..577Z} is computed. If significant peaks other than the planetary signal (i.e., contaminating peaks, typically induced by stellar rotation or its harmonics) are present, \texttt{MINERVA} attempts an iterative pre-whitening, evaluating the intruder peaks in decreasing order of power. The highest-power intruder is temporarily fitted and subtracted, and the periodogram of the residuals is evaluated: if the significance of the planetary signal degrades (e.g., the power drops significantly), the intruder is rejected as a likely alias of the planet. Conversely, if the planetary signal improves, the intruder peak is assumed to be the stellar rotation period, triggering a simultaneous fit of both the planetary and stellar signals.

Alternatively, if the planetary signal is the dominant peak and no valid intruders are identified, the code performs a reverse pre-whitening. A preliminary Keplerian model of the planet is subtracted from the raw data, and a GLS periodogram is computed on the residuals to search for underlying stellar activity signatures. If a significant stellar peak emerges, it triggers a simultaneous fit; otherwise, the pipeline proceeds with a planet-only model. 

Depending on the outcome of these steps, the algorithm fits the unbinned raw RV series with either a planet-only Keplerian model or a simultaneous multi-component model (planet plus stellar activity). The formal modelling within the \texttt{MINERVA} pipeline is executed through a frequentist, non-linear least-squares minimization approach rather than a stochastic Bayesian sampling (MCMC), ensuring the high computational efficiency required to simulate thousands of independent observing campaigns. The optimization is performed using the Trust Region Reflective (\texttt{trf}) algorithm implemented via \texttt{scipy.optimize.curve\_fit} \citep{Virtanen2020SciPy-NMeth}, which allows for robust convergence under bounded constraints (e.g., imposing $K_{\text{pl}} \ge 0$). 

To break the strong degeneracies between overlapping signals, the orbital period $P_{\text{pl}}$ and the transit mid-point $T_0$ of the planet are kept fixed to the known photometric transit ephemerides. Similarly, when a simultaneous fit is triggered, the stellar rotation period $P_{\text{star}}$ is fixed to the exact frequency of the peak identified by the upstream GLS periodogram analysis. Consequently, the free parameters optimized by the minimizer are exclusively the planetary RV semi-amplitude $K_{\text{pl}}$, and a triplet of local parameters for each independent observing season $s$: the seasonal zero-point offset $\gamma_{\rm s}$, and the Fourier sine/cosine amplitudes $A_{\rm s}$ and $B_{\rm s}$. This seasonal parametrization allows the model to independently capture the amplitude evolution and phase shifts driven by the dynamic evolution of active regions from one year to the next.

Following the minimization, the pipeline evaluates the goodness of fit by computing the reduced chi-squared metric, $\chi^2_{\text{red}}$. To ensure realistic and conservative uncertainties on the fitted parameters, the formal geometric uncertainty on the amplitude ($\sigma_{\text{K, formal}}$) is adjusted to a final value ($\sigma_{\text{K}} = \sigma_{\text{K, formal}} \times \max(1.0, \sqrt{\chi^2_{\text{red}}})$), which directly penalizes the signal-to-noise ratio ($K_{\text{fit}}/\sigma_{\text{K}}$) in case of under-weighted residuals. Finally, model selection is performed via the Bayesian information criterion ($\text{BIC} = \chi^2 + k \ln N$), evaluating the statistical evidence of the full model against the null hypothesis.

We define a signal as fully and robustly recovered only when the following criteria are simultaneously satisfied by the final fit:
\begin{itemize}
\item[-] Statistical significance: The adopted model (planet or simultaneous) must be strongly statistically favoured over a no-planet model. This is quantified by a Bayesian information criterion difference ($\Delta\text{BIC}$) $> 5$ \citep{jeffreys1998theory}.
\item[-] Physical accuracy: The recovered parameters must be consistent with the injected inputs. Specifically, the fitted RV semi-amplitude ($K_{\rm fit}$) must match the injected planetary semi-amplitude within a $3\sigma_{\rm K}$ confidence interval. If a simultaneous fit is performed, we do not constrain the fitted stellar semi-amplitude, allowing the model to freely adapt to the intrinsic variability of the active regions. However, the recovered stellar period must strictly match the true rotational period ($P_{\rm rot}$), or its first two harmonics ($P_{\rm rot}/2$ and $P_{\rm rot}/3$), within a 15\% tolerance.
\item[-] Required precision: The signal-to-noise ratio ($K/\sigma_{\rm K}$) must exceed a user-defined threshold, ensuring that the planetary mass can be determined with the precision necessary for subsequent atmospheric characterisation.
\end{itemize}
If the final fit does not meet all the above criteria, an additional set of five RV measurements is added, and the entire process is repeated until the threshold is reached or the full dataset is exhausted.

\section{Results and discussion}
\label{sec:results}
\subsection{Code validation}
\label{sec:validation}
\begin{table*}[htbp]
    \centering
    \caption{Sample used to test the code.}
    \renewcommand{\arraystretch}{1.3}
    % Usa \textwidth per occupare tutta la pagina (utile per table*)
    \resizebox{\textwidth}{!}{%
    \begin{tabular}{l c c c c c c c c c c}																	
         \hline																	
         \hline
         \rule{0pt}{2.8ex} Name & Sp. Type & $V$ (mag) & $M_\star$ ($M_{\odot}$) & $M_{\rm p}$ ($M_{\rm Jup}$) & $P$ (d) & $\log R'_{\rm HK}$ & $N_{\rm RV, literature}$& \multicolumn{2}{c}{$N_{\rm RV}$$_{\texttt{MINERVA}}$} & Succ. (\%) \\
         \cline{9-10}
         \rule{0pt}{2.8ex} & & & & & & & & (All) & (Success) & \\
         \hline																	
         HAT-P-24b$^{(a)}$  & F8\,V   & 11.82 & $1.19 \pm 0.04$ & $0.69 \pm 0.03$       & $3.355240 \pm 0.000007$             & -4.7$^(i)$  & 18         & $15^{+16}_{-5}$ & $14 \pm 5$   & 85 \\
         TOI-132b$^{(b)}$   & G8\,V   & 11.29 & $0.97 \pm 0.06$ & $22.4 \pm 1.9$        & $2.1097019^{+0.000012}_{-0.000011}$ & -5.02 $\pm$ 0.13     & 20         & $40^{+1}_{-16}$ & $27 \pm 8$   & 50 \\
         K2-370b$^{(c)}$    & G5\,V   & 11.11 & $0.96 \pm 0.04$ & $0.035 \pm 0.004$     & $2.14084 \pm 0.00004$               & -4.49 $\pm$ 0.03     & 136        & $270$           & $219 \pm 78$ & 10 \\
         Corot-10b$^{(d)}$  & K1\,V   & 15.22 & $0.89 \pm 0.05$ & $2.75 \pm 0.16$       & $13.2406 \pm 0.0002$                & -4.4$^(i)$  & 19$^{(j)}$& $10$            & $11 \pm 3$   & 100 \\
         GJ 625b$^{(e)}$    & M2\,V   & 10.17 & $0.30 \pm 0.07$ & $0.0089 \pm 0.0016$   & $14.628^{+0.012}_{-0.013}$          & -5.5 $\pm$ 0.2      & 140        & $148^{+133}_{-73}$ & $129 \pm 58$ & 81 \\
         GJ 720Ab$^{(f)}$   & M0.5\,V & 9.84  & $0.57 \pm 0.06$ & $0.043\pm 0.002$      & $19.466 \pm 0.005$                  & -5.03 $\pm$ 0.04     & 138        & $260^{+1}_{-90}$   & $171 \pm 36$ & 50 \\
         GJ 3942b$^{(g)}$   & M0\,V   & 10.25 & $0.63 \pm 0.07$ & $0.0225 \pm 0.0019$   & $6.91 \pm 0.04$                     & -4.55 $\pm$ 0.05     & 142        & $265^{+25}_{-65}$  & $209 \pm 52$ & 59 \\
         GJ 3998c$^{(h)}$   & M1\,V   & 10.83 & $0.50 \pm 0.05$ & $0.020^{+0.003}_{-0.002}$& $13.740 \pm 0.016$               & -5.01$^(i)$     & 136        & $195^{+75}_{-61}$  & $171 \pm 47$ & 69 \\
         \hline																	
    \end{tabular}%
    }																	
    \vspace{0.5em}																	
    \par																	
    \captionsetup{width=\textwidth}																	
    \caption*{\textbf{Note.} $^{(i)}$ $\log R'_{\rm HK}$ values not available in the literature were computed using \texttt{MINERVA}. $^{(j)}$ Of the 19 observations acquired, the authors reported that 7 were affected by moonlight contamination. The $N_{\rm RV_{\rm MINERVA}}$ (All) and (Success) columns report the full simulation set (median with 16$^{\rm th}$/84$^{\rm th}$ percentiles) and the successful recoveries only (mean $\pm$ standard deviation), respectively. Values without uncertainty bounds in the $N_{\rm RV_{\rm MINERVA}}$ (All) column indicate highly peaked distributions where percentiles perfectly coincide (e.g., because the vast majority of simulations hit the saturation limit, or because the planetary signal is unequivocally strong). $^{(a)}$ \citet{Kipping2010ApJ...725.2017K}; $^{(b)}$ \citet{Diaz2020MNRAS.493..973D}; $^{(c)}$ \citet{Sozzetti2024MNRAS.535..531S}; $^{(d)}$ \citet{Bonomo2010A&A...520A..65B}; $^{(e)}$ \citet{SuarezMascareno2017A&A...605A..92S}; $^{(f)}$ \citet{GonzalezAlvarez2021A&A...649A.157G}; $^{(g)}$ \citet{Perger2017A&A...608A..63P}; $^{(h)}$ \citet{Affer2016A&A...593A.117A}.}																	
    \label{table:sampletest}																	
\end{table*}
In this section, we describe the procedure adopted to validate the performance of \texttt{MINERVA}. The primary objective of the code is to estimate the number of RV measurements ($N_{\rm RV}$) required to determine the planetary masses with a user-defined precision. For validation purpose, we set this target precision to match the uncertainties reported in the literature. It is important to note from the outset that our approach yields a deliberately conservative estimate of the required observational effort. Since \texttt{MINERVA} does not employ sophisticated post-processing techniques (such as Gaussian processes) to model and mitigate correlated stellar signals, methods commonly used in recent discovery papers, our predicted $N_{\rm RV}$ represents a robust upper limit. 

We constructed a test catalogue of stars spanning spectral types from F to M, encompassing a wide range of magnitudes and activity levels, to ensure the method's robustness across different stellar characteristics. These targets host planets whose masses have been previously constrained using high-resolution RV data, predominantly from HARPS-N. Our goal is to compare the $N_{\rm RV}$ estimated by \texttt{MINERVA} with the actual number of observations used in the discovery papers. The targets comprising the test sample are listed in Table \ref{table:sampletest}.		

We configured our code to reproduce the observational conditions described in the original studies. Specifically, we adopted the exact temporal sampling, and used the stellar and planetary parameters, generally well constrained in the literature, to generate the RV time series (e.g., stellar mass, age, rotation period, $\log R'_{\rm HK}$ index, and planetary RV semi-amplitude).
For each system, we simulated 100 independent time series and evaluated the distribution of the $N_{\rm RV}$ required to recover the planetary mass at the reported literature precision.
To ensure a robust recovery of the planetary signal and to account for statistical fluctuations arising from the generation of the stochastic noise floor and activity patterns, we systematically extended the original time series baselines for all targets by adding further observing epochs consistent with the initial cadence. Furthermore, since the stellar activity parameters were not always well constrained in the literature, we performed the test under two different assumptions: allowing \texttt{MINERVA} to generate these parameters automatically via its internal empirical relations, and fixing them to the available literature values. Both approaches yielded consistent results, thereby supporting the reliability of the empirical relations implemented within the code.

To evaluate the overall observational effort estimated by \texttt{MINERVA}, we compared our results with the literature data. For our selected sample, the cumulative sum of RV measurements reported in the discovery papers is 749 points. It is crucial to note that this literature value inherently represents a sample of pure successes: discovery papers are published precisely because the planet was successfully detected, creating a natural publication bias. In reality, observational campaigns that encounter highly unfavourable stochastic conditions, such as epochs where the stellar spot distribution drastically degrades the planetary signal, are often halted or result in non-detections. Therefore, to make a direct comparison with the literature, we first calculated the required effort by considering exclusively our successful simulations, effectively filtering out the extreme worst-case scenarios. Under these favourable conditions, the \texttt{MINERVA} distributions accurately reproduce the observational effort. A target-by-target inspection reveals that the actual number of RVs reported in the literature consistently falls within the $1\sigma$ standard deviation of our success-only means (Table \ref{table:sampletest}). For instance, the literature reports 136 observations for GJ 3998\,b (vs. our $171 \pm 47$), 140 for GJ 625\,b (vs. $129 \pm 58$), 18 for HAT-P-24\,b (vs. $14 \pm 5$), and 20 for TOI-132\,b (vs. $27 \pm 8$). Summing these means across the sample yields a total of $951 \pm 126$ observations, showing a compatibility within 2$\sigma$ with the literature baseline of 749 points. 
The slight expected overestimation is due to the highly conservative, unsupervised statistical thresholds imposed by our pipeline (e.g., a strict $\Delta$BIC $>5$), compared to the heavily supervised analyses used in discovery papers.

A further confirmation comes from analysing the full distributions, which include the right-censored ``extreme'' cases. We find that the literature values systematically align with the lower tail (i.e., the 16th percentile) of the \texttt{MINERVA} full distributions. This indicates that our pipeline correctly identifies the absolute minimum $N_{\rm RV}$ required to claim a detection under optimal stochastic conditions. This lower bound perfectly mirrors the behaviour of actual observing campaigns, which are typically halted as soon as the minimal detection threshold is achieved.

However, while the lower percentiles and the success-only metric validates the core algorithm against past discoveries, relying on them to plan a future blind survey would be risky. A real, automated survey cannot guarantee that it will only encounter favourable conditions. Therefore, to provide a truly robust and realistic estimate for future survey planning, we must account for the full probability distribution. Because the median is not an additive operator, we computed the full-sample statistics using a Monte Carlo bootstrap approach, drawing $10^4$ random realisations from the complete simulation outcomes of all targets. Using this statistical treatment, the median total observational effort for the entire sample rises to $1145 \pm 130$ RV measurements. This transition highlights an important methodological implication: while literature discoveries ($749$ points) often represent best-case samplings of the underlying probability distribution, the $1145$ points estimated by the \texttt{MINERVA} bootstrap represent a conservative upper limit for the cost of a blind survey. This metric guarantees a robust planetary recovery even under the worst-case stochastic of stellar red noise, acting as a safe observational threshold that modern post-processing analyses can subsequently optimise.

The significant discrepancy concerns K2-370\,b, where the recovery procedure with \texttt{MINERVA} remained challenging even after extending the time series. For this target, the failure rate remained high, with 90\% of the simulations ending in non-detections. This is inherently linked to the extreme level of stellar activity of this target. Across the 100 simulations, \texttt{MINERVA} generated an RV semi-amplitude due to stellar activity of $\rm \sim15\,m\,s^{-1}$, significantly overshadowing the Keplerian signal amplitude of $\rm 5.7\,m\,s^{-1}$. 
While the ratio between activity and planetary amplitudes is comparable to some of the M-dwarf cases (which also required a large number of observations), K2-370 falls into the extremely active regime ($\log R'_{\rm HK} > -4.5$). Consequently, the simulated stellar activity signal exhibits a highly stochastic behaviour, higher than the true signal. This complexity hampers the analytic pre-whitening and simultaneous fitting of the activity contribution, making the detection of the planetary signal exceedingly difficult. This interpretation is perfectly consistent with the findings of \citet{Sozzetti2024MNRAS.535..531S}, who were similarly unable to detect the planetary signal through classical periodogram analysis and were forced to adopt a more sophisticated approach using gaussian process regression. 

Excluding this extreme case, the $N_{\rm RV}$ estimated by \texttt{MINERVA} is consistent with the observational effort reported in the literature. This demonstrates that \texttt{MINERVA} provides a highly reliable, albeit appropriately conservative, prediction of the observing time required to achieve a specified precision in planetary mass determination, even without incorporating advanced, computationally expensive post-processing techniques like gaussian process regression.

Since \texttt{MINERVA} is primarily designed as a forward-modelling tool for population studies, evaluating individual literature targets with fixed nominal parameters represents a simplified baseline for code validation. However, to preserve the inherent probabilistic nature of \texttt{MINERVA} and assess the impact of observational uncertainties on single targets, we tested the performance of the stellar parameter bootstrapping module described in Sect. \ref{subsection:starpop_properties}. By resampling physical parameters (such as stellar mass, effective temperature, and $\log R'_{HK}$) within their reported literature uncertainties ($\pm\sigma$), we found that for successful signal recoveries, the required $N_{\rm RV}$ remains remarkably stable, particularly for planet-dominated systems. Conversely, propagating parameter uncertainties primarily affects the overall recovery rate (\% Success) in active or borderline signal-to-noise ratio regimes, where fluctuations in derived stellar activity levels can occasionally mask the planetary signal within the allocated baseline. A remarkable exception is K2-370b: while under fixed nominal parameters its elevated activity level results in a low recovery rate (10\%), the parameter bootstrapping allows the pipeline to sample slightly lower activity states within the literature uncertainties, driving the detection efficiency up to 41\% for successful realizations. This demonstrates that while nominal literature values accurately reproduce historical discovery baselines, the parameter bootstrapping option provides users with a valuable diagnostic tool to evaluate undersampling risks in active stellar environments.

While our baseline validation focuses on circular orbits, and the parameter bootstrapping module addresses uncertainties in stellar physics, another critical unknown in planning blind RV surveys is the planetary orbital architecture. \texttt{MINERVA}'s ability to inject user-defined eccentricities and arguments of periastron provides a powerful tool for telescope time allocation proposals, allowing users to stress-test their scheduling strategies against unfavourable orbital configurations. To evaluate the impact of unmodeled eccentricity on our retrieval module, we performed dedicated test runs varying $e$ from 0 to 0.7, fixing $\omega = 0^\circ$ to simulate a highly asymmetric, worst-case RV curve.

As expected, the required number of observations generally increases with eccentricity, since the spectral power of the signal leaks from the fundamental frequency into higher harmonics, dampening the main GLS peak. For small to moderate eccentricities ($e \le 0.3$), the median $N_{\rm RV}$ for successful detections increases but typically remains statistically consistent within $1\sigma$ of the circular baseline. At higher eccentricities ($e \ge 0.5$), the fraction of non-detections rises significantly. Physically, a non-detection implies that the maximum number of RV points allocated for the simulated campaign was exhausted before the dampened fundamental peak could accumulate enough statistical significance to cross the detection thresholds. However, we note that the failure threshold is highly target-dependent: depending on the stellar noise properties and the intrinsic signal-to-noise ratio, the circular retrieval algorithm can sometimes successfully recover signals even up to $e=0.7$, albeit at the cost of a substantially larger observational effort.

Ultimately, the primary advantage of evaluating the impact of eccentricity within \texttt{MINERVA} lies in its predictive power for survey design. By simulating worst-case orbital scenarios, observers can accurately quantify the hidden overhead in RV epochs required to secure a detection. This ensures that telescope time proposals are built with sufficient observational margins, preventing premature campaign failures due to unmodeled eccentricities and maximizing the scientific yield of expensive ground-based follow-up campaigns.

\subsection{The \textit{Ariel} sample}
\label{sec:arielsample}
In this section, we consider targets listed in the \textit{Ariel} catalogues. Ultimately, the mission will observe the final MRS, which is currently still under development. Therefore, our analysis is based on the preparatory MCS, which serves as the overarching pool of potential targets. The MCS is composed of two main sub-catalogues: the \textit{Ariel} Known Planet Catalogue (hereafter KPC), which includes confirmed transiting planets with mass estimates available for most of them; and the \textit{Ariel} TESS Planet Catalogue (hereafter TPCs catalogue), which lists TESS Objects of Interest (TOIs). These TOIs have not yet been validated but, if confirmed, would represent promising \textit{Ariel} targets and be subsequently promoted to the KPC. 

As shown in \citet{DiMaio2023A&A...669A.150D}, achieving accurate atmospheric retrievals requires a planetary mass precision of approximately 40\% for giant planets, where a primordial atmosphere can be assumed, and around 30\% for smaller planets with potential secondary atmospheres. Based on these recommendations, we included in our sample all KPC targets with planetary mass uncertainties exceeding 40\%, as well as all targets in the TPCs catalogue.
We note that the inclusion of the TPCs catalogue, which contains targets that have not yet been validated, may in principle include false positives. However, since our work primarily aims at a statistical characterisation of the sample, the presence of potential false positives does not  affect our analysis.
\begin{figure}[h!]
	\centering
         {\subfloat[\label{fig:CRSsample_mvTeff}]{\includegraphics[width=0.5\textwidth]{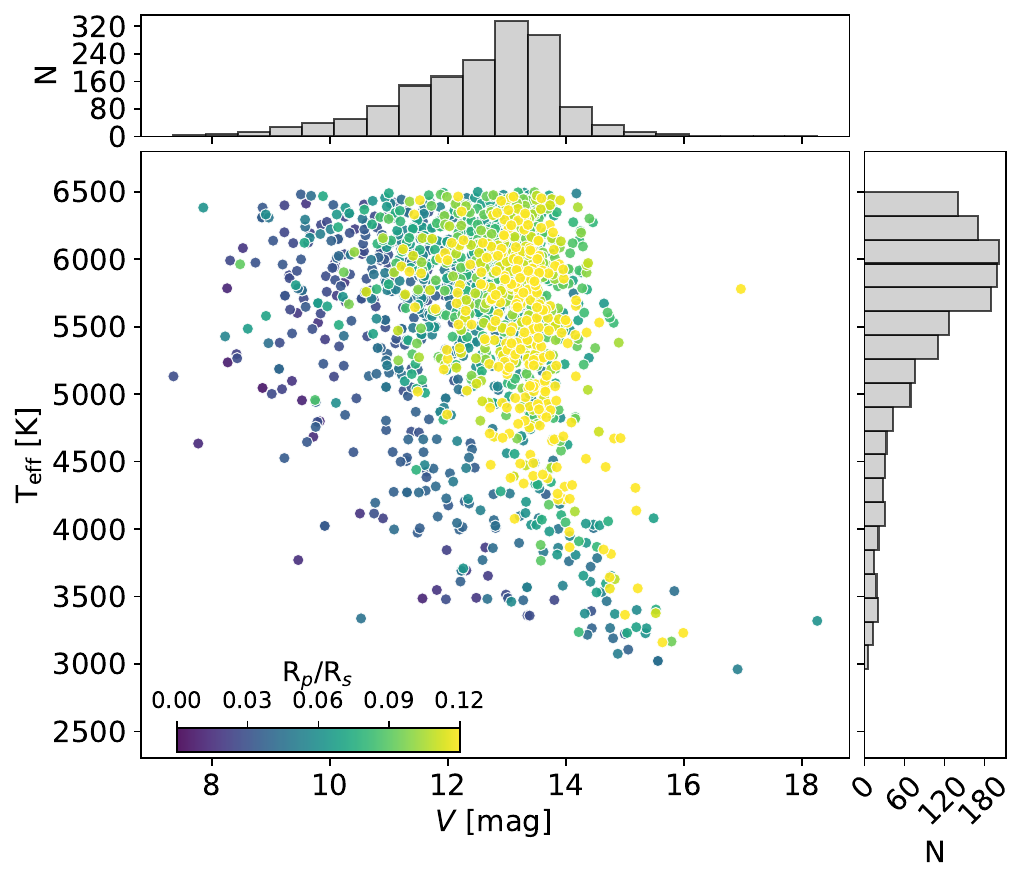}} \\
    \subfloat[\label{fig:CRSsample_periodradius}]{\includegraphics[width=0.5\textwidth]{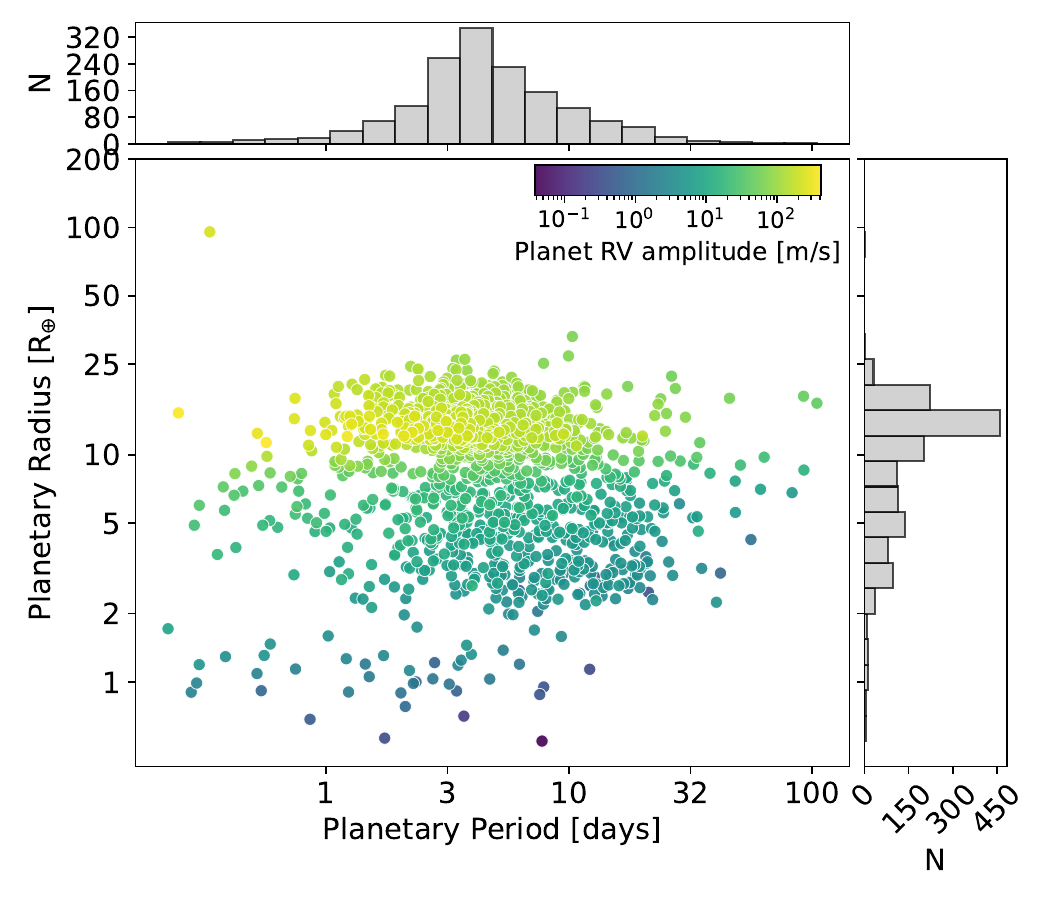}}
    }	\caption[]{Overview of the \textit{Ariel} sample used in our simulations. (a) Visual magnitude vs. effective temperature for each target star, colour-coded according to the planet-to-star radius ratio ($R_{\rm p}/R_{\rm s}$). (b) Planetary radius as a function of orbital period for the same sample, with each point colour-coded according to the expected RV semi-amplitude of the planetary signal. The attached marginal histograms in both panels show the corresponding 1D distributions.}
	\label{fig:CRSsample}
\end{figure}

 Using the versions of these catalogues released in July 2025, we identified a final sample of 1541 targets out of 2854 available. These targets were restricted to main-sequence FGKM stars with effective temperatures below 6500 K and masses below 1.4 $M_{\odot}$, consistent with \texttt{MINERVA}’s design specifications.
 This final sample includes both single and multi-planet systems, with 91 targets originating from the KPC. 

Figure \ref{fig:CRSsample} provides a visual overview of the stellar and planetary properties within our sample. In particular, 492 targets in our sample have planetary radii smaller than 7\,$R_\oplus$; for these objects, precise mass determination is especially critical to enable robust atmospheric characterization in preparation for \textit{Ariel}.

\subsubsection{\texttt{MINERVA} applied to the \textit{Ariel} subsample}
\begin{figure}[h!]
	\centering
         {\includegraphics[width=0.5\textwidth]{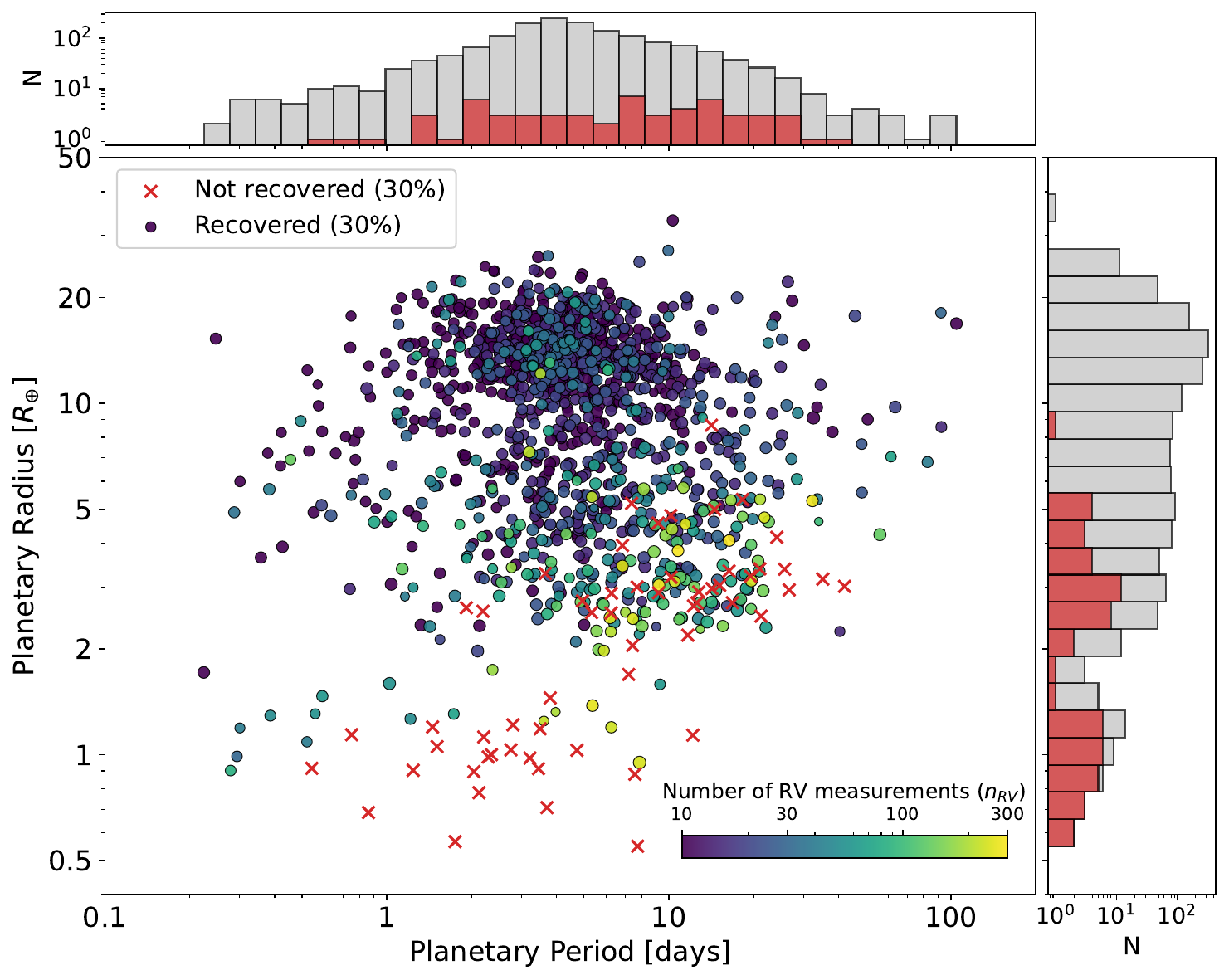}} \\
    \caption[]{Planet radius versus orbital period for planets in our sample. Planets with a recovered mass from \texttt{MINERVA} are shown as filled circles, colour-coded by number of RVs ($N_{\rm RV}$), whereas planets without a mass estimate are indicated by red crosses. The top and right marginal histograms display the 1D distributions of the orbital period and planetary radius, respectively, comparing the recovered (grey) and unrecovered (red) populations. Both main axes are in logarithmic scale. The inset colour bar indicates the number of RV observations required to recover the planetary mass with a 30\% precision.}
	\label{fig:period_radius_recovered}
\end{figure}
For each target in our sample, we simulated a RV time series assuming a HARPS-N–like instrumental precision. 
Following the procedure described in Sect.~\ref{subsec:time}, the temporal sampling and the maximum baseline were tailored to the planetary orbital period. For the bulk of the sample, we generated up to 300 data points adopting the HD~164922 reference cadence. Conversely, for USP planets ($P < 2$ days), we applied the intensive TOI-1807 scheduling, consisting of up to 160 high-cadence measurements. We emphasize that the temporal samplings implemented in \texttt{MINERVA} reflect realistic, multi-target observing campaigns governed by standard telescope time-sharing policies. While target-specific phase-optimization algorithms \citep[e.g., \texttt{KOBEsim};][]{Balsalobre-Ruza2023A&A...669A..18B} can potentially reduce the required number of measurements for individual high-priority systems by scheduling epochs at optimal orbital phases, they demand flexible queue operations and  precise prior ephemerides. By employing standard unoptimized cadences, our approach instead provides a conservative baseline for planning large-scale surveys under typical operational constraints. 
For all the systems, we required a precision of 30\% in the planetary mass determination. The stellar activity level for each target was estimated based on its rotation period ($P_{\rm rot}$), as described in Sect. \ref{subsec:activity_sim}. Following the framework in Sect. \ref{subsection:starpop_properties}, these properties were derived from stellar age estimates by assuming our sample is representative of the solar neighbourhood with a constant stellar birthrate.

\begin{figure}[h!]
	\centering
	\includegraphics[width=0.5\textwidth]{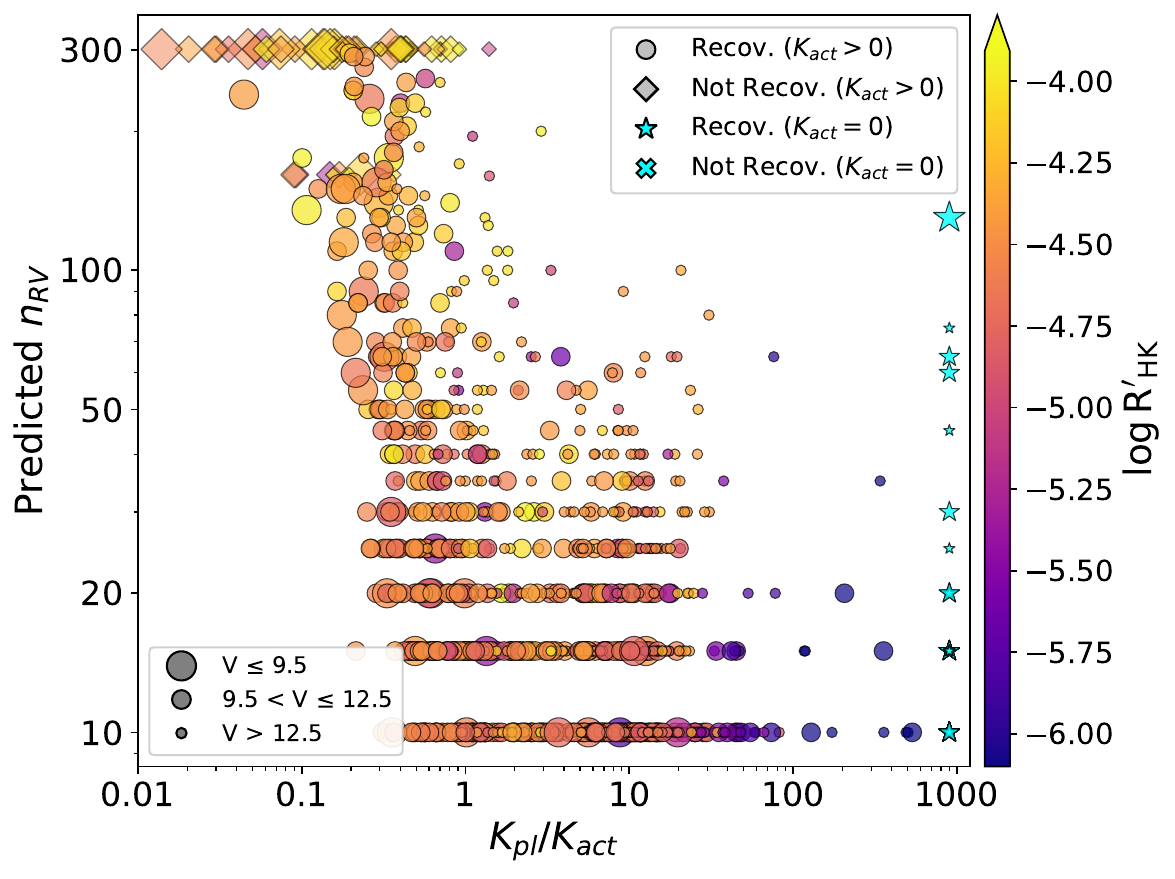}
	\caption{Predicted number of RV observations required for a 30\% mass precision as a function of the planetary-to-activity amplitude ratio ($K_{\rm pl}/K_{\rm act}$). Marker colour maps the stellar activity index ($\log R'_{\rm HK}$), while size scales with the host star's V magnitude. Shapes denote simulation outcomes: circles (success) and diamonds (non-detection) for stars with modelled activity ($K_{\rm act} > 0$). Targets orbiting stars without a convective envelope ($K_{\rm act} = 0$) are highlighted in cyan using star symbols (success) and crosses (non-detection); these are assigned a $K_{\rm pl}/K_{\rm act} = 9 \times 10^2$ for display purposes.}
\label{fig:predictedRV}
\end{figure} 

According to our results, for 1482 out of 1541 planets we could successfully recover the mass in less then 300 RV,  including 67 planets from the KPC.  Here, successful recovery means that the planetary mass was determined with a precision better than 30\% within the allocated observational baseline. Within the sample of 1482 recovered planets, 435 have radii smaller than 7\,R$_\oplus$ and masses below 39 M$_\oplus$. 

Figures \ref{fig:period_radius_recovered} and \ref{fig:predictedRV} illustrate the prediction of the \texttt{MINERVA} pipeline across the parameter space of the \textit{Ariel} catalogues. Specifically, Fig. \ref{fig:period_radius_recovered} shows the planetary radius versus the orbital period for both the KP and TPCs samples, highlighting the successfully recovered planets (colour-coded by the required number of RVs) alongside the non-recovered targets (red crosses). As clearly shown by the projected histograms, the non-recovered systems predominantly cluster in the regime of small planetary radii ($R_{\rm p} \lesssim 4-5 \, R_\oplus$) across a wide range of orbital periods, highlighting the parameter space where the Keplerian semi-amplitude drops below the noise threshold.
For the successfully recovered targets, the colour gradient suggests  that the observational effort increases drastically for smaller and wider-orbiting planets. Specifically, while larger planets ($R_{\rm p} \gtrsim 5 \, R_\oplus$) with short orbital periods ($P < 10$ days) typically require a limited effort of $\sim 10 - 50$ RV measurements, the cost rises for smaller or more distant targets. For planets with $R_{\rm p} \lesssim 5 \, R_\oplus$ or those with periods exceeding $\sim 6$ days, the required effort frequently exceeds $100 - 200$ observations.

Figure \ref{fig:predictedRV} provides a deeper insight into the physical drivers of this observational cost, plotting the predicted $N_{\rm RV}$ required to achieve a 30\% mass precision as a function of the planetary-to-activity amplitude ratio ($K_{\text{pl}}/K_{\text{act}}$). A general  trend is observed: targets where the Keplerian signal strongly dominates over the stellar activity ($K_{\text{pl}}/K_{\text{act}} \gg 1$) systematically converge toward lower observational costs, requiring down to 10–20 epochs. Conversely, as $K_{\text{pl}}/K_{\text{act}}$ drops below unity, the required number of observations increases. This effect culminates in the non-recovered targets (diamonds), which accumulate on the top-left of the plot and form a distinct horizontal sequence saturating at our simulation ceiling ($N_{\text{RV}} = 300$, or 160 for USP planets).
\begin{table}[ht]\footnotesize																	
    \centering																	
    \caption{Summary of the recovered planets and the number of RVs and nights required to obtain a 30\% significance for the whole sample and for the small planet sample.}																	
    \begin{tabular}{c c c c c}																	
         \hline																	
         \hline	
         \multicolumn{5}{c}{\rule{0pt}{2.8ex}Whole sample}\\[1ex]
         \hline
         \rule{0pt}{2.8ex}Magnitude range (mag) & $t_{\rm exp}$(s) & N. targets & N. RVs & N. nights \\[1ex]
         \hline																	
         $V \leq 8.5$             & 600  & 10   & 1688   & 46.89 \\  
         $8.5 < V \leq 9.5$   & 900  & 34   & 4084   & 151.26 \\
         $9.5 < V \leq 10.5$  & 900  & 78  & 5142  & 190.44 \\
         $10.5 < V \leq 11.5$ & 1200 & 187  & 9317  & 431.34 \\
         $11.5 < V \leq 12.5$ & 1800 & 322  & 12284  & 796.19 \\
         $12.5 < V \leq 13.5$ & 2700 & 602  & 13006  & 1204.26 \\
         $V \geq 13.5$            & 3600 & 308  & 10902  & 1312.28 \\
         \hline
         \rule{0pt}{2.8ex}{Total}  & {...} & {1541} & {56423} & {4132.66} \\[1ex]
         \hline	
         \multicolumn{5}{c}{\rule{0pt}{2.8ex}Small planet sample}\\[1ex]
         \hline
         \rule{0pt}{2.8ex}Magnitude range (mag) & $t_{\rm exp}$(s) & N. targets & N. RVs & N. nights \\[1ex]
         \hline
         $V \leq 8.5$             & 600  & 7   & 1488 & 41.33 \\  
         $8.5 <  V \leq 9.5$   & 900  & 20  & 3223 & 119.37 \\
         $9.5 <  V \leq 10.5$  & 900  & 34  & 3365 & 124.63 \\
         $10.5 < V \leq 11.5$ & 1200 & 58  & 5371 & 248.66 \\
         $11.5 < V \leq 12.5$ & 1800 & 42  & 5552 & 359.85 \\
         $12.5 < V \leq 13.5$ & 2700 & 24  & 2861 & 264.91 \\
         $V \geq 13.5$            & 3600 & 41  & 4960 & 597.04 \\ 
         \hline
         \rule{0pt}{2.8ex}{Total}  & {...} & {226} & {26820} & {1755.79} \\[1ex]
         \hline
    \end{tabular}																	
\vspace{0.5em}																	
\par																	
\captionsetup{width=1\textwidth}					\label{table:RVminerva}																	
\end{table}		

The colour-coding, mapping the $\log R'_{\rm HK}$, reveals the physical root of this distribution. The unrecovered targets (diamonds at $K_{\text{pl}}/K_{\text{act}} \lesssim 0.3$) are predominantly highly active stars (bright yellow and orange markers, $\log R'_{\rm HK} \gtrsim -4.5$), demonstrating how intense stellar activity modulations effectively mask low-amplitude Keplerian signals within the allowed observational baseline. In contrast, quiescent targets (dark purple markers, $\log R'_{\rm HK} \lesssim -5.0$) populate the regime of higher amplitude ratios, requiring significantly fewer observations for a successful mass recovery.
On the far right of the plot, targets lacking a convective envelope ($K_{\text{act}} = 0$) are represented by cyan star symbols (successful recoveries) and cyan crosses (non-detections), arbitrarily placed at $K_{\text{pl}}/K_{\text{act}} = 9 \times 10^2$ for display purposes. Because no activity-driven RV modulations from spots or plages are expected for these stars, no stellar activity signal was injected ($K_{\text{act}} = 0$). Consequently, their required observational effort is dictated strictly by the balance between $K_{\text{pl}}$ and $\delta RV_i$. While most of these inactive targets converge quickly at low observational costs ($N_{\text{RV}} \lesssim 30$--$50$), they span a broad range of required epochs, extending up to $\sim 140$ observations for a single outlier system. This high observational cost demonstrates that even in the complete absence of stellar activity, a very small planetary semi-amplitude $K_{\text{pl}}$ combined with a fainter host star (higher photon noise $\sigma_{\text{RV,ph}}$) yields a low detection Signal-to-Noise Ratio ($K_{\text{pl}} / \delta RV_{\rm i}$), requiring a significantly larger dataset to disentangle the Keplerian signal from pure white noise and achieve the target 30\% mass precision.
Finally, the marker size, proportional to the host star's visual magnitude ($V$), indicates that while apparent brightness influences the baseline photon noise, with fainter stars generally requiring slightly more observations at a fixed amplitude ratio, the competition between the planetary signal and the stellar activity amplitude ($K_{\text{pl}}/K_{\text{act}}$) remains the dominant physical driver of the total observing cost.

\begin{figure}[ht!]
	\centering
	\includegraphics[width=0.5\textwidth]{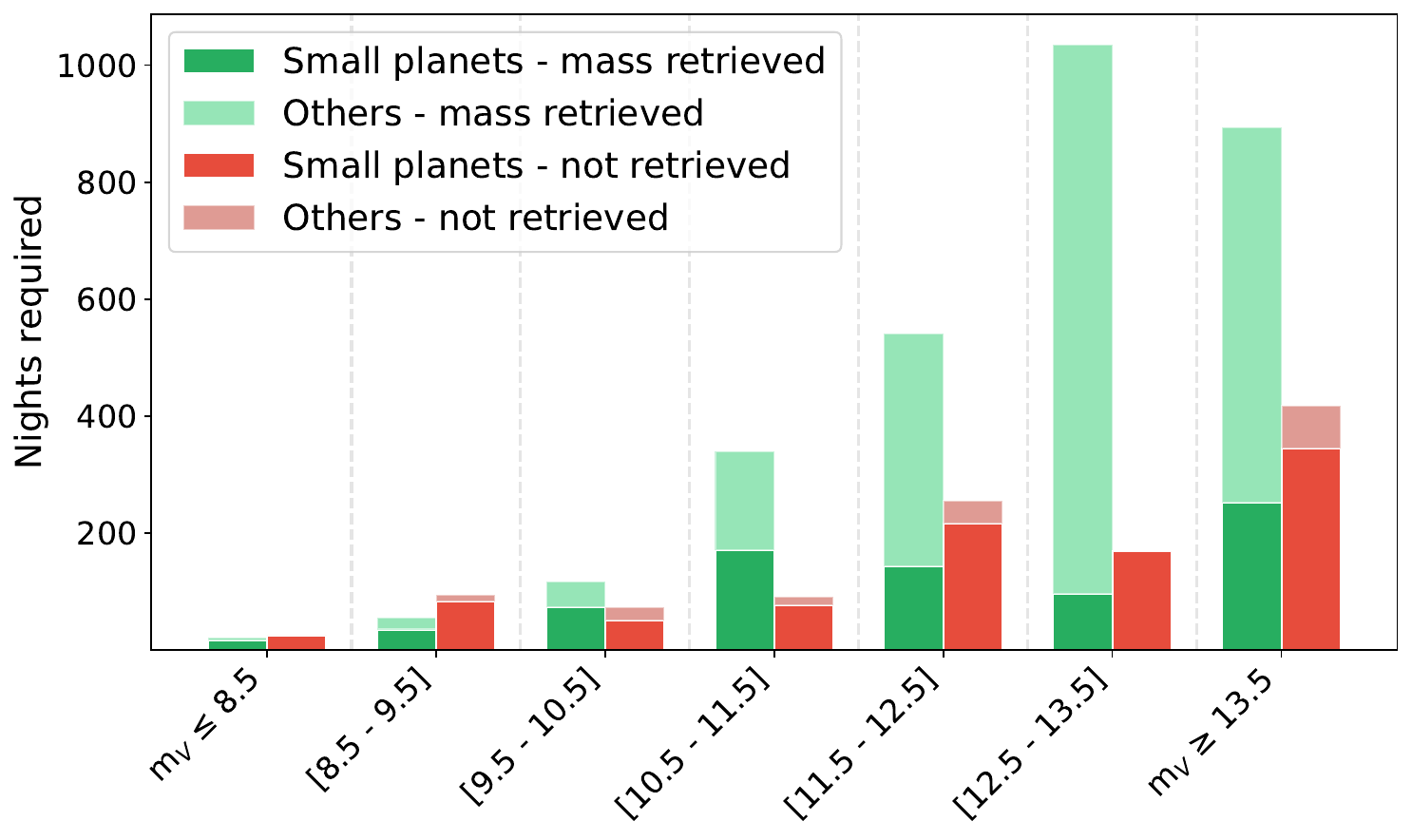}
	\caption{Total observing time (in nights) per visual magnitude ($V$) bin, categorised by mass retrieval status. Green bars indicate successfully recovered targets, while red bars represent unrecovered systems that reached the maximum observational baseline (160 RVs for USPs, 300 for others) without achieving the required mass precision. Within each bar, darker shades denote the small planet sub-sample ($R_{\rm p} < 4 R_\oplus$), while lighter shades represent the rest of the catalogue.}
\label{fig:hist_nights}
\end{figure}

To translate the required number of RV epochs into actual observing time, we simulated a campaign using a high-resolution spectrograph mounted on a 4-m class telescope, analogous to HARPS-N at the TNG. For each target, we adopted standard, fixed exposure time bins as a function of visual magnitude ($V$), representative of realistic observing strategies (T$_{\rm exp}$ column in Table \ref{table:RVminerva}). 
We assumed a standard overhead of 5 minutes per exposure, accounting for telescope slewing, target acquisition, instrument setup, and detector readout, and considered an average observing night of 9 hours, in agreement with the standard guidelines for TNG time allocation proposals.
 We note that these estimates represent net clear-sky time budgets; observers translating these numbers into actual observing proposals should account for site-specific seasonal weather losses.
The resulting time estimates, organised as a function of visual magnitude, are summarised in Table \ref{table:RVminerva}. 	

We evaluated the observing time for two scenarios: the entire sample (1541 targets) and a sub-sample of 226 targets with $R_{\rm p} < 4 R_\oplus$ (hereafter, the small planet sample). 
When the entire sample is considered, a total of 4133 nights is required to achieve the necessary mass precision. If we restrict our focus solely to the small planet sample, the required time amounts to 1756 nights (Fig. \ref{fig:hist_nights}). Notably, while small planets make up only 15\% of the targets, they consume roughly 42\% of the total observing time.

In total, 59 targets were not recovered, an overwhelming 86\% of which (51 targets) belong to the small-planet regime ($R_{\rm p} < 4\,R_\oplus$). Analysis of these non-detections shows that an unfavourable signal-to-noise ratio is the universal limiting factor: the unrecovered targets exhibit an amplitude ratio $K_{\rm pl}/K_{\rm act} < 1.4$, with 58 of them falling below unity. This demonstrates that for the vast majority of the unrecovered sample, the planetary signal is entirely buried within the stellar jitter, preventing a successful retrieval with the methodology described in this work regardless of whether the failure is driven by extreme stellar activity ($\log R'_{\rm HK} > -4$, for 11 targets) or by low-amplitude planetary signal (48 targets). Even for these unrecovered cases, the time investment was accounted for by considering the full observational effort expended before the simulation reached its saturation ceiling (i.e., the maximum number of RV measurements allowed by the specific cadence). 
\begin{table}[t]
    \centering
    \caption{Summary of the simulated samples and the requested amount of RV epochs estimated by \texttt{MINERVA}.}
    % Questo comando aumenta lo spazio verticale di tutte le righe del 30%
    \renewcommand{\arraystretch}{1.3}
    \resizebox{\columnwidth}{!}{%
    \begin{tabular}{l c c c c}												
         \hline															
         \hline	
         % --- PRIMA SEZIONE: OVERVIEW DEI CAMPIONI ---
         \multicolumn{5}{c}{\rule{0pt}{3ex}{Sample overview}} \\[1ex]
         \hline
         \rule{0pt}{2.8ex} Sample & Criteria & Targets & Unres. & Total nights\\[1ex]
         \hline
         Entire sample & All $R_{\rm p}$, All $V$ & 1541 & 59 & 4133 (3007$^{a}$) \\
         Small planets & $R_{\rm p} < 4 R_\oplus$, All $V$  & 226 & 51 & 1795 (789$^{a}$) \\ 
         Bright small & $R_{\rm p} < 4 R_\oplus$, $V \le 12.5$ & 161 & 34 & 894 (442$^{a}$) \\
         \hline
    \end{tabular}%
    }
    \vspace{0.5em}															
    \par																	
    \captionsetup{width=\columnwidth}
    {\textbf{Note.}} $(^{a})$ Values in parentheses indicate the observing nights calculated by considering only the successfully recovered targets, excluding the hidden cost of unrecovered ones.
    \label{table:Summary}		
\end{table}
In a real-world scenario, such an extensive allocation reflects the total resources a blind survey would consume to ensure robustness against stellar activity, even when certain targets remain unrecovered. However, if we exclude these non-recovered cases, which are clearly dominated by elevated stellar activity, the required observational effort drops drastically (by $\sim$ 27\%). Considering only the successfully recovered systems, the total budget reduces to 3007 nights. Within this idealised scenario, characterising the small planet sample would require 789 nights (a reduction of $\sim 55\%)$, thereby consuming more than a quarter ($\sim 26\%$) of the entire successful night budget.  A comprehensive comparison of the
required observing nights is summarised in Table \ref{table:Summary}.

While the majority of the overall sample requires fewer than 20 RVs per target, the observational effort is heavily driven by the most challenging systems. Specifically, 208 targets in our simulation require more than 50 observations each, with small planets accounting for nearly two-thirds of this intensive subset (132 planets, or 63\%). If we focus exclusively on successfully recovered systems, 149 targets require more than 50 observations to reach the 30\% mass precision threshold; notably, 81 of these (54\%) belong to the small planet sample. Given this substantial time investment, characterising the smallest planets with 4-m class facilities is highly demanding. For a more efficient observational campaign, leveraging larger aperture telescopes would be necessary to significantly reduce exposure times or the number of RV measurements. 	

From the KP and TPCs catalogues, the final MRS of the \textit{Ariel} mission will be extracted, comprising about 1000 targets that will be observed to reach the mission's objectives. There are several possible choices to define the MRS, and the results of our simulations can contribute to the selection and prioritization process. %Here, we discuss two possible scenarios. 
Some scenarios are discussed in the Appendix \ref{app:Scenarios}.

Our results highlight the critical importance of dedicated, targeted observing campaigns, such as the ArMS programme at the TNG. Although the launch is still five years away, starting these efforts early is absolutely crucial. As demonstrated by our simulations, characterising the masses of small planets, especially those orbiting active stars, is a highly time-consuming endeavour that requires a sustained, multi-year observational investment. Crucially, securing these mass measurements well ahead of launch directly informs the mission's core observing strategy. Accurate mass constraints are a fundamental prerequisite to evaluate the expected atmospheric signals, calculate the precise number of transits required for each target, and ultimately drive the final selection and prioritization of the MRS. This is precisely why the \textit{Ariel} Mass Determination Working Group is actively seeking coordination among the various independent observing programs, with the ultimate goal of minimising duplications and optimising both telescope time and community resources.

\section{Conclusions}\label{sec:conclusions}Precise measurements of planetary mass are a prerequisite for the accurate retrieval of atmospheric properties. As demonstrated in previous atmospheric retrieval studies \citep[e.g.,][]{Batalha2019ApJ...885L..25B, Changeat2020ApJ...896..107C, DiMaio2023A&A...669A.150D}, a mass precision better than $\sim$30\% is essential to break the degeneracies between mass, cloud coverage, and chemical composition, particularly for super-Earths and sub-Neptunes with secondary atmospheres. To address the challenge of planning the extensive RV follow-up campaigns required for upcoming missions, we developed \texttt{MINERVA} (MInimum observations for accurate Exoplanetary mass estimation with Radial Velocity Analysis). This tool simulates realistic RV time series to provide robust estimate of the observing effort required to recover signals of transiting planets.

\texttt{MINERVA} incorporates a comprehensive noise framework that accounts for realistic stellar activity, actual observation scheduling, and robust signal recovery. A defining feature of the tool is its highly modular architecture: every functional block, from the stellar activity engine and the observational scheduler to the signal recovery pipeline, is designed to be fully customizable. This modularity allows the user to easily modify individual components, adapt the code to different instrument specifications, or incorporate more advanced stellar models as they become available.

Specifically, we implemented a flexible activity injection system based on the $\log R'_{\rm HK}$ index that models time-correlated signals, ranging from simple sinusoids to chaotic, multi-spot variability, providing conservative estimates for active, young stars. Furthermore, the simulations adopt realistic observing cadences based on actual GAPS/HARPS-N campaigns, incorporating seasonal gaps and visibility constraints,  while an incremental retrieval module, combining Generalised Lomb-Scargle periodograms with robust orbital fitting and Bayesian model comparison ($\Delta$BIC), determines the exact number of epochs needed to satisfy strict detection criteria and precision requirements.

Prior to its application to future surveys, we rigorously validated \texttt{MINERVA} against a test sample of known planetary systems spanning F to M spectral types. By mirroring the observational conditions and temporal samplings of the original discovery papers, we demonstrated that our pipeline accurately reproduces the real-world observational effort. When accounting for the inherent publication bias of successful literature discoveries, \texttt{MINERVA}’s estimates ($951 \pm 126$ RVs) aligned closely with the 749 RVs reported in the literature. Moreover, a Monte Carlo bootstrap analysis established a median total effort of $1145 \pm 130$ RVs for the full distribution. These validation results confirm that \texttt{MINERVA} provides a reliable, deliberately conservative upper limit for blind surveys, identifying safe observational thresholds that guarantee planetary recovery even without the use of computationally expensive post-processing techniques such as Gaussian process regression.

Building on this validation framework, the inclusion of user-defined orbital eccentricity and argument of periastron allows the community to realistically stress-test observational campaigns against unfavourable, non-circular configurations. Our tests highlight that while the fast GLS-based circular retrieval remains robust for small to moderate eccentricities, highly eccentric targets demand significantly larger RV allocations and carry inherently higher non-detection risks. For such extreme regimes, \texttt{MINERVA} provides the essential forward-modelling tool to quantify the exact hidden overhead of assuming a circular scheduling baseline.

We then applied the tool to the \textit{Ariel} candidate catalogues, focussing specifically on the challenging small-planet regime ($R_{\rm p} < 4 \, R_\oplus$). Our analysis reveals that neglecting the complex activity patterns of younger targets and the intrinsic faintness of the host stars leads to a significant underestimation of the telescope time needed. 

Furthermore, our simulations allowed us to define a quantitative selection criterion to optimise future RV follow-up campaigns. By analysing the targets that failed to reach the required mass precision within the allocated baseline, we demonstrated that non-detections are systematically driven by a critical threshold in the signal-to-jitter ratios ($K_{\rm pl}/K_{\rm act} \lesssim 1$). This unfavourable ratio arises either from extreme stellar activity ($\log R'_{\rm HK} > -4$), where the jitter level dominates the time series, or from the intrinsic smallness of the planetary signal compared to the stellar noise floor.  Since these challenging configurations account for the entirety of our unrecovered cases, an overwhelming 86\% of which belong to the small-planet regime, we propose a selection framework for finalising the \textit{Ariel} MRS, whereby targets falling within these specific physical regimes, where high stellar activity or severe photon noise drastically increase the observational demand, must either be prioritised for immediate, intensive, and long-term monitoring, or be explicitly excluded from the follow-up sample to prevent the severe depletion of the community's observational budget.

Crucially, these results highlight the strategic importance of ongoing efforts to measure planetary masses, such as the dedicated ArMS large programme at the TNG. Although the \textit{Ariel} launch is still five years away, starting these preparatory efforts immediately is of paramount importance. Measuring the masses of small planets, especially in the presence of stellar activity, is an inherently time-consuming process that requires a long baseline and sustained observational investment. However, securing these mass measurements well ahead of launch is not merely a matter of time management; it directly informs the mission's core observing strategy.
 Accurate mass constraints are a fundamental prerequisite to evaluate the expected atmospheric scale heights, calculate the precise number of transits required for each target, and ultimately drive the final selection and prioritization of the MRS. Waiting until launch to conduct these observations would severely compromise the ability to select the optimal targets for Tier 2 and Tier 3 observations. This necessity for long-term planning is also the primary driver behind the \textit{Ariel} ``Mass determination'' Working Group, which is actively seeking coordination among the various international observing programs. In this context, \texttt{MINERVA} serves as a valuable tool to optimise both time and resources, ensuring that the community can effectively allocate the observational budget required to unlock the full scientific potential of the \textit{Ariel} mission.

%__________________________________________________________________

\begin{acknowledgements}
We thank the anonymous referee for their insightful comments and constructive suggestions, which significantly improved the quality of this manuscript.
The authors acknowledge the support of the ASI-INAF agreement 2021-5-HH.0, the INAF Guest Observer Grant (Normal) ``ArMS: the \textit{Ariel} Masses Survey Large Program at the TNG'', the project ``INAF-Astrofisica Fondamentale GAPS2'', INAF GO Large Grant 2023 GAPS-2 and the INAF Mini-Grant ``Impact of planetary Masses and Radii Estimates on the Atmospheric retrievals - IMaREA'', according to the INAF Fundamental Astrophysics funding scheme. The research activities described in this paper were carried out with contribution of the Next Generation EU funds within the National Recovery and Resilience Plan (PNRR), Mission 4 - Education and Research, Component 2 - From Research to Business (M4C2), Investment Line 3.1 - Strengthening and creation of Research Infrastructures, Project IR0000034 – ``STILES - Strengthening the Italian Leadership in ELT and SKA''. This work has been carried out with support and data from the Global Architecture of Planetary Systems (GAPS) Project.
\end{acknowledgements}

\balance{
    \bibliographystyle{aa}
    \bibliography{bibliography}}

%-------------------------------------------------------------------

\appendix

\onecolumn

\section{\texttt{MINERVA} setup: additional figures}\label{app:MINERVAsetup}
\begin{figure*}[h!]
	\centering
\includegraphics[width=\textwidth]{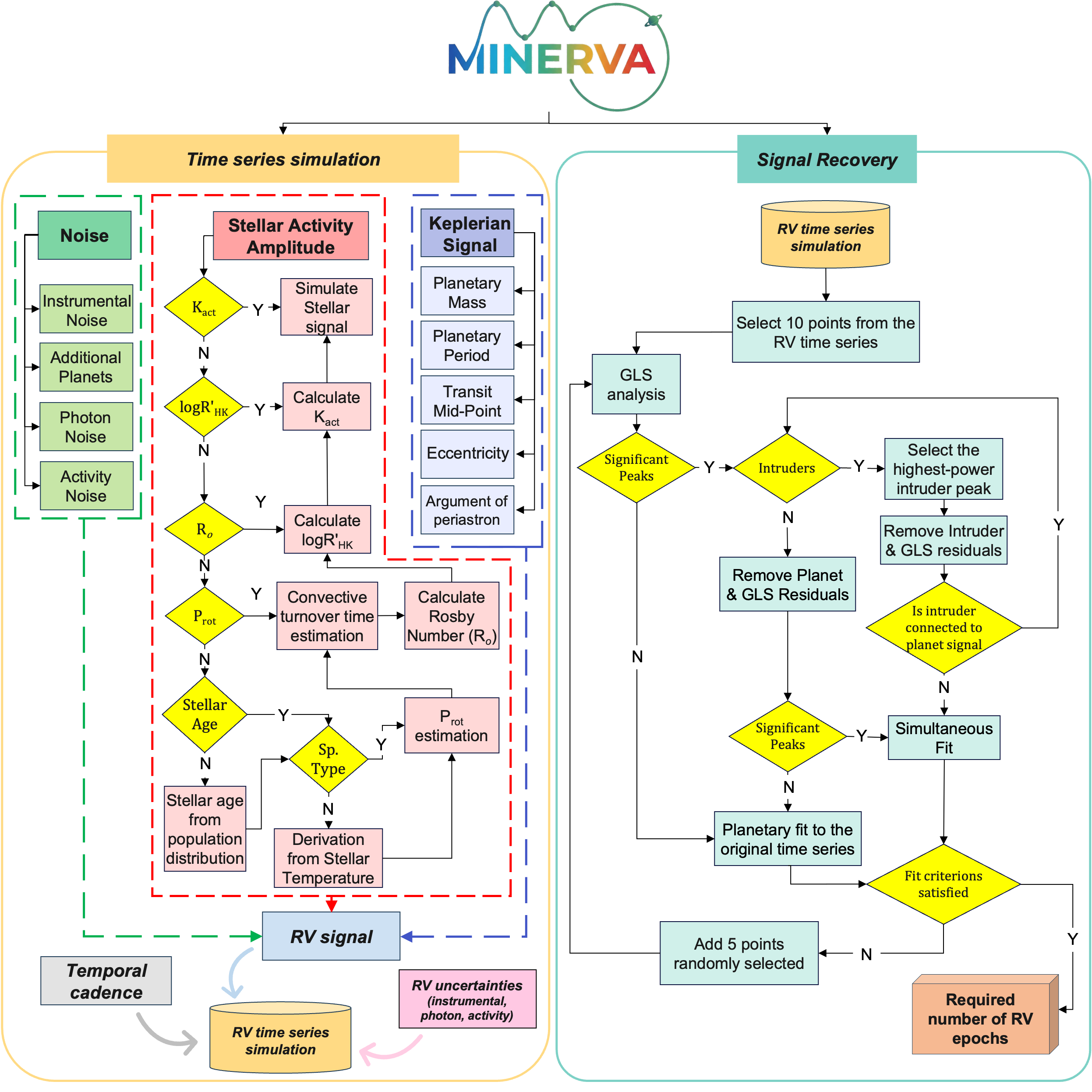}
	\caption{Flowchart of \texttt{MINERVA}, illustrating its two main stages: simulation and recovery. The simulation (left branch) is based on a temporal cadence (Sect. \ref{subsec:time}) and combines three components: the noise contribution (green dashed box, Sect. \ref{subsec:noise}), the stellar activity signal (red dashed box), with its amplitude defined as described in Sect. \ref{subsection:starpop_properties}, and the Keplerian signal (blue dashed box, Sect. \ref{subsec:planetary_signal}). The planetary recovery procedure (right branch) is detailed in Sect. \ref{subsec:recovery}.
    }
\label{fig:diagramma}
\end{figure*} 

\twocolumn
\section{Evaluation of possible scenarios for the \textit{Ariel} targets selection and prioritization}\label{app:Scenarios}
Focussing on the smallest planets ($R_{\text{p}} < 4\,R_\oplus$), for which accurate knowledge of the mass is essential for reliable atmospheric characterization, the \textit{Ariel} KPC currently includes about 100 targets with a mass uncertainty below 30\%, which are expected to be incorporated into the final MRS. 
However, relying solely on this existing sample would be insufficient to fulfill \textit{Ariel}'s core statistical objectives. The small-planet regime is known to be highly diverse in terms of bulk compositions and atmospheric properties. To perform a robust population study and identify significant demographic trends, the mission requires a larger, statistically significant sample. 

Assuming the MRS will dedicate roughly 20\% of its observing slots (i.e., $\sim$ 200 targets) to this critical planetary regime to achieve these demographic goals, we estimate that $\sim$ 100 additional small planets ($R_{\text{p}} < 4\,R_\oplus$) could be selected from the recovered TESS candidate sample (the TPCs catalogue analysed in Sect.\ref{sec:arielsample}) and fully characterised.

\begin{figure}[h!]
	\centering
	\includegraphics[width=0.5\textwidth]{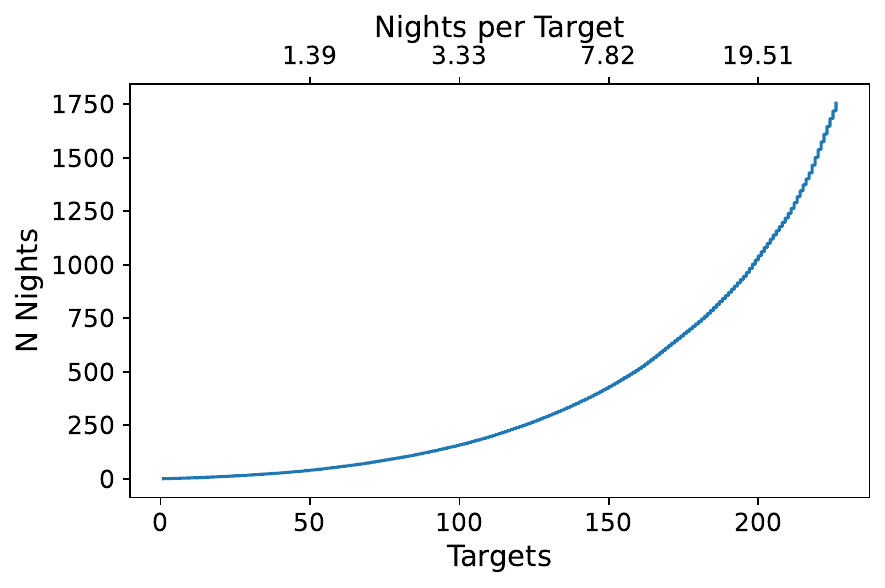}
	\caption{Cumulative distribution of required observation nights for the small planet sample. The Y-axis shows the total number of nights needed to observe a given number of targets. Targets are sorted by increasing observation cost. The top X-axis indicates the cost in nights for the individual target at that position in the sequence.}
\label{fig:cumulative_distribution}
\end{figure} 

To evaluate the observational effort required to finalise this selection of these $\sim$ 100 additional small planets, we defined and compare two different prioritization strategies:

\begin{itemize}
\item {Cost-optimised selection}: This approach prioritises purely the observational efficiency. We select the 100 additional planets that require the absolute minimum amount of observing time, based on the cumulative distribution (Fig. \ref{fig:cumulative_distribution}). 
\item {Demographically-balanced selection}: This approach forces a uniform demographic coverage across the small-planet radius range ($R_{\rm p} < 4\,R_\oplus$). We divide the sample into bins of 0.5\,$R_\oplus$ (starting from $R_{\rm p} < 1\,R_\oplus$). Crucially, to minimise the total effort while maintaining this uniform coverage, we still apply a cost-optimisation criterion within each bin, selecting only the most observationally favourable targets available for that specific radius range.
\end{itemize}

We evaluated both strategies on two parent pools: the unrestricted small-planet sample ($226$ targets) and a magnitude-limited sub-sample ($m_{\text{V}} \le 12.5$, $161$ targets). The brightness threshold reflects the practical operational limits of current 4-m class high-resolution spectrographs. The detailed comparison of required observing nights, success-only budgets, and demographic coverage across all scenarios is summarized in Table~\ref{table:StrategySummary}.

\begin{table}[htbp]
    \centering
    \caption{Detailed summary of observing costs and $R_{\rm p} < 2\,R_\oplus$ coverage for each investigated selection strategy.}
    % Questo comando aumenta lo spazio verticale di tutte le righe del 30%
    \renewcommand{\arraystretch}{1.3}
    \resizebox{\columnwidth}{!}{%
    \begin{tabular}{l c c c c}												
         \hline															
         \hline
         % --- SECONDA SEZIONE: STRATEGIE DI SELEZIONE ---
         \multicolumn{5}{c}{\rule{0pt}{4ex}{Selection strategies}} \\[1ex]
         \hline
         \rule{0pt}{2.8ex} Strategy & Condition & Sel. targets & Nights & Cov. $R_{\rm p} \!<\! 2\,R_\oplus$\\[1ex]
         \hline
         Cost-opt. & Unrestr. $V$ & 100 & 159 & Poor \\
         Demo-bal. & Unrestr. $V$  & 103 (101$^{a}$) & 563 (307$^{a}$) & Excellent \\ 
         Cost-opt. & $V \le 12.5$ & 100 & 198 & Poor \\
         Demo-bal. & $V \le 12.5$ & 100 (100$^{a}$) & 376 (275$^{a}$) & Excellent\\ 
         \hline	
    \end{tabular}%
    }
    \vspace{0.5em}															
    \par																	
    \captionsetup{width=\columnwidth}
    {\textbf{Note.}} $(^{a})$ Values in parentheses indicate the observing nights calculated by considering only the successfully recovered targets, excluding the hidden cost of unrecovered ones.
    \label{table:StrategySummary}		
\end{table}

Comparing these strategies reveals a trade-off between observational efficiency and demographic completeness:
\begin{itemize}
    \item {Bias of the cost-optimised strategy (both samples):} Regardless of whether we use the unrestricted or the magnitude-limited pool, this approach inherently biases the sample toward larger sub-Neptunes ($R_{\text{p}} \gtrsim 2\,R_\oplus$). This is visually represented by the region below the red horizontal line in Fig. \ref{fig:required_nights_small}, which encloses the most time-efficient targets but highlights the severe under-representation of ultra-small planets ($R_{\text{p}} < 2\,R_\oplus$).

    \begin{figure}[h!]
	\centering
	\includegraphics[width=0.5\textwidth]{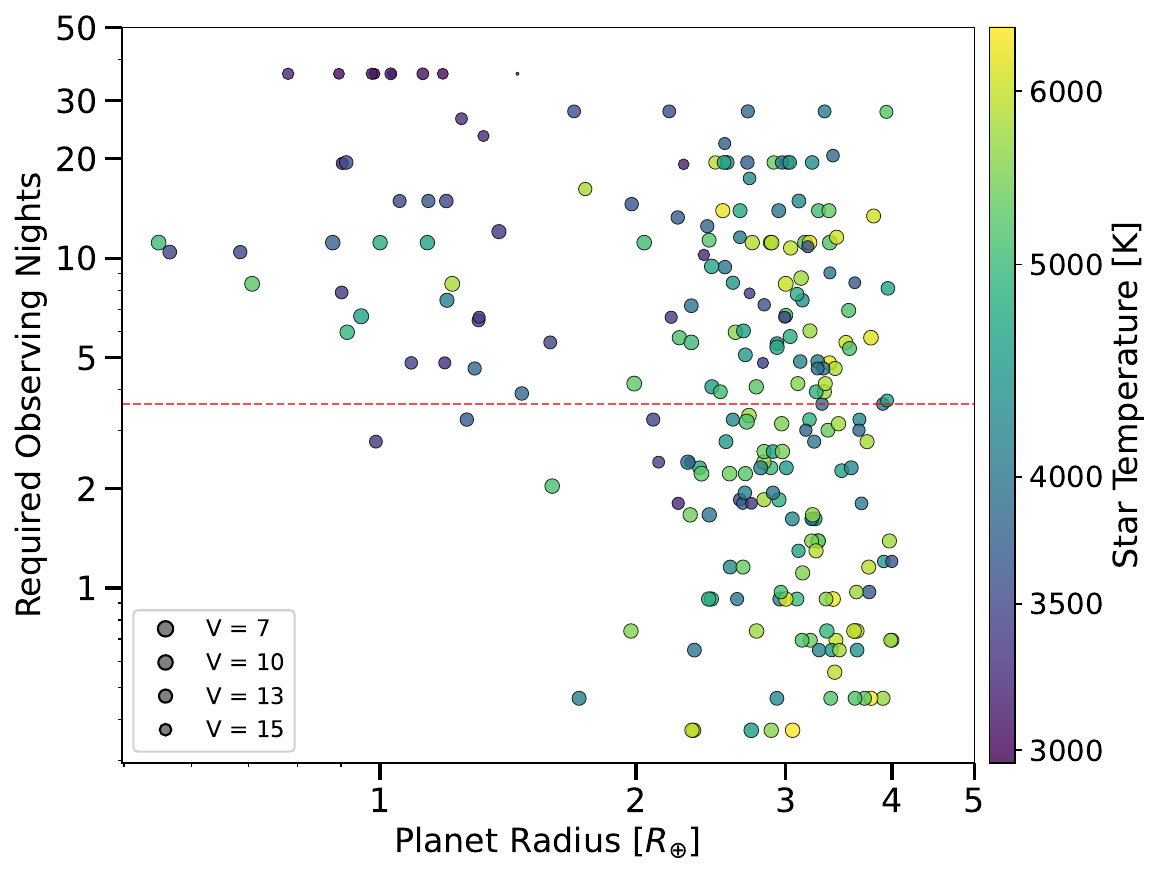}
	\caption{Predicted required nights to reach 30\% mass precision for small planets in our sample (log-log scale). Marker colour maps the stellar effective temperature, and size is inversely proportional to visual magnitude ($V$). The red horizontal line denotes a threshold of $\sim 8$ nights, enclosing the 100 most time-efficient targets (requiring $< 4$ nights each).}
\label{fig:required_nights_small}
\end{figure} 

    \item {The 'hidden cost' penalty (unrestricted sample):} When applied to the full, unrestricted pool, the demographically-balanced strategy incurs a massive time penalty (a factor of $\sim 3.5$ higher cost than the cost-optimised baseline). This is driven by faint, active targets that consume telescope time until they hit the simulation saturation limits without achieving the $30\%$ mass precision.

    \item {Impact of the magnitude cut ($m_{\text{V}} \le 12.5$):} Restricting the pool to bright targets halves the overall observing budget and reduces the balanced strategy's cost multiplier from $\sim3.5$ down to $\sim1.9$ (Table~\ref{table:StrategySummary}). However, as shown by the vertical jumps at the rightmost boundaries of Fig. \ref{fig:cumulative_RV_bright}, a 'hidden cost' persists: even within this bright regime, a fraction of challenging targets still hits the saturation limits ($160$ RVs for USPs, $300$ RVs for standard periods) without a successful mass recovery.

\begin{figure}[h!]
	\centering
	\includegraphics[width=0.45\textwidth]{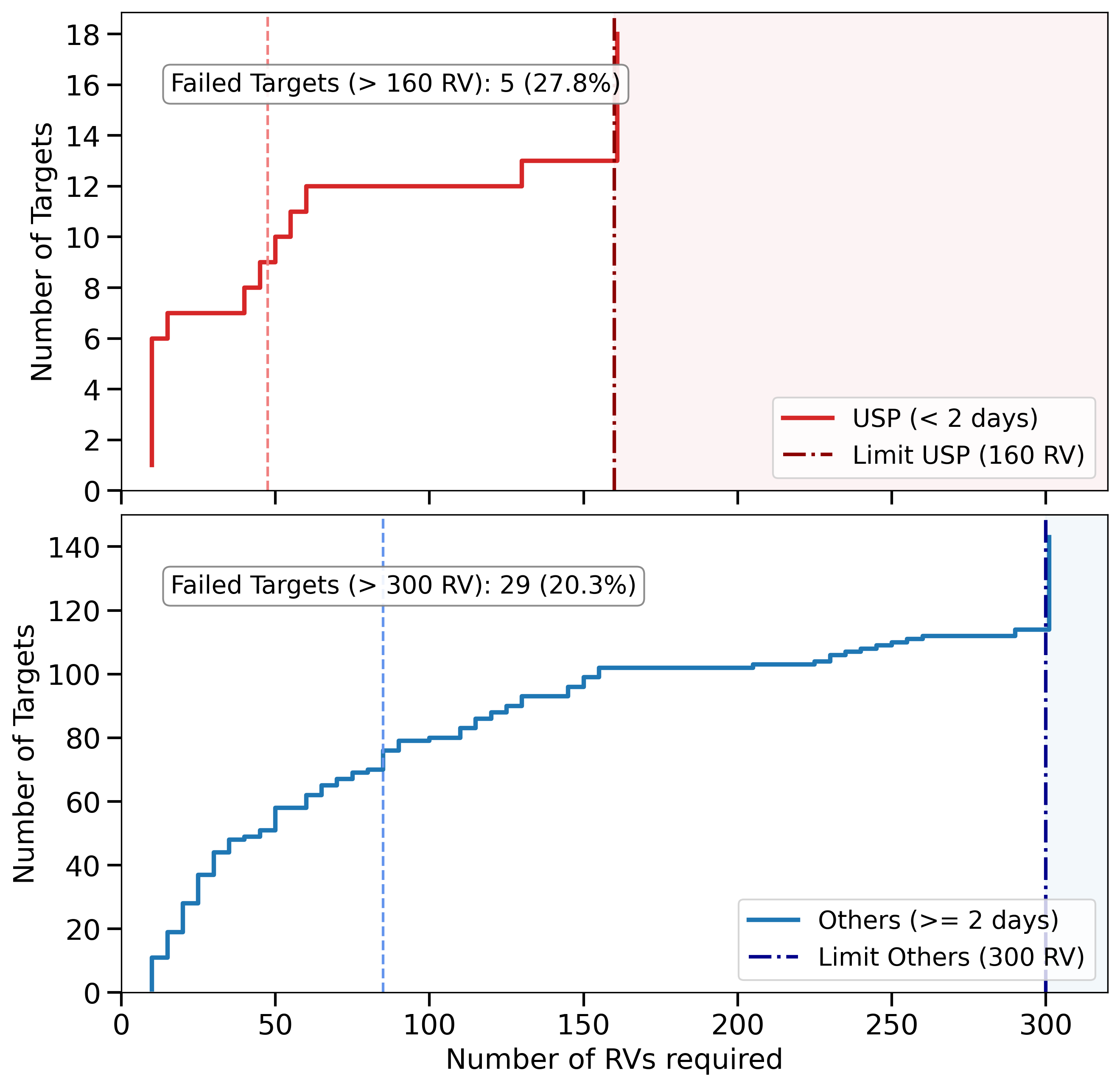}
	\caption{Cumulative number of targets vs. predicted RV epochs required for 30\% mass precision, focussing on bright small-planet sub-sample ($V \le 12.5$, $R_{\rm p} < 4\,R_\oplus$). The top and bottom panels display UPS ($P < 2$ days) and standard ($P \ge 2$ days) planets, respectively. Vertical dashed lines indicate the median RVs for successfully recovered systems. Thick dash-dotted lines mark the simulation saturation limits (160 and 300 RVs). The final vertical steps at these boundaries represent the non-recovered targets, which account for 27.8\% and 20.3\% of their respective populations.}
\label{fig:cumulative_RV_bright}
\end{figure}

    \item {The success-only scenario (balanced strategy):} Attempting to avoid these saturated targets by strictly selecting only successfully recovered planets reduces the observing time (values in parentheses in Table \ref{table:StrategySummary}). However, this idealised approach severely degrades the demographic coverage, drastically depleting the ultra-small regime ($R_{\text{p}} < 1\,R_\oplus$) to just a few viable targets.
\end{itemize}

As a final remark, we note that the mass measurements of the small planets can be obtained both before the \textit{Ariel} launch, currently scheduled for 2031, and during the mission's operations. Based on our estimates for the magnitude-limited sample ($V \le 12.5$), allocating approximately 30 nights per year on 4-m class telescopes from now until launch would yield a total of roughly 150 observing nights. This dedicated effort would be sufficient to cover a substantial fraction of the required baseline: approximately 40\% of the baseline for the demographically-balanced strategy (150 out of 376 nights), or an impressive 76\% if the cost-optimised approach is adopted (150 out of 198 nights).

Furthermore, we stress that our simulated time budget represents a highly conservative upper limit. To evaluate the total theoretical cost, we computed the required baseline assuming zero prior knowledge of the planetary masses. In reality, the targets included in the KPC already possess baseline mass measurements (albeit sometimes with uncertainties exceeding the 30\% threshold), and a significant fraction of the TESS candidates populating the TPCs catalogue is currently being actively monitored by various teams using multiple facilities. While it is challenging to accurately quantify the total number of telescope nights currently devoted worldwide to the mass determination of small TESS candidates, these ongoing community efforts will inevitably reduce the actual observing time required to finalise the \textit{Ariel} MRS.

\end{document}